\documentclass[manuscript,screen]{acmart}  
\setcopyright{ieeecopyright}
\copyrightyear{2023}
\usepackage{afterpage}
\usepackage{soul}  

\makeatletter
\g@addto@macro{\UrlBreaks}{\UrlOrds}
\makeatother

\usepackage{subcaption}
\usepackage{url}
\usepackage{listings}       
\usepackage{xcolor}
\definecolor{backcolour}{RGB}{250, 250, 250}   %
\definecolor{codegreen}{RGB}{16, 124, 2}       
\definecolor{codepurple}{RGB}{170, 0, 217}     
\definecolor{codered}{RGB}{154, 0, 18}         
\lstdefinestyle{gcolabstyle}{
  basicstyle=\ttfamily\small,
  backgroundcolor=\color{backcolour},   
  commentstyle=\itshape\color{codegreen},
  keywordstyle=\color{codepurple},
  stringstyle=\color{codered},
  numberstyle=\ttfamily\footnotesize\color{darkgray}, 
  breakatwhitespace=false,         
  breaklines=true,                 
  captionpos=b,                    
  keepspaces=true,                 
  numbers=left,                    
  numbersep=5pt,                  
  showspaces=false,                
  showstringspaces=false,
  showtabs=false,                  
  tabsize=2
}
\newcommand{\urim}[1]{\href{https://web.archive.org/web/#1/https://www.cnn.com/}{#1}}

\begin{document}

\title[Right HTML, Wrong JSON]{Right HTML, Wrong JSON: Challenges in Replaying Archived Webpages Built with Client-Side Rendering}

\author{Michele C. Weigle}
\orcid{0000-0002-2787-7166}
\affiliation{%
  \institution{Old Dominion University}
  \city{Norfolk}
  \state{VA}
  \country{USA}
  \postcode{23529}
}
\email{mweigle@cs.odu.edu}

\author{Michael L. Nelson}
\orcid{0000-0003-3749-8116}
\affiliation{%
  \institution{Old Dominion University}
  \city{Norfolk}
  \state{VA}
  \country{USA}
  \postcode{23529}
}
\email{mln@cs.odu.edu}

\author{Sawood Alam}
\orcid{0000-0002-8267-3326}
\affiliation{%
  \institution{Internet Archive}
  \city{San Francisco}
  \state{CA}
  \country{USA}
  \postcode{94118}
}
\email{sawood@archive.org}

\author{Mark Graham}
\orcid{}
\affiliation{%
  \institution{Internet Archive}
  \city{San Francisco}
  \state{CA}
  \country{USA}
  \postcode{94118}
}
\email{mark@archive.org}

\renewcommand{\shortauthors}{Weigle, Nelson, Alam, and Graham}

\begin{abstract}
Many web sites are transitioning how they construct their pages.  The conventional model is where the content is embedded \emph{server-side} in the HTML and returned to the client in an HTTP response.  Increasingly, sites are moving to a model where the initial HTTP response contains only an HTML skeleton plus JavaScript that makes API calls to a variety of servers for the content (typically in JSON format), and then builds out the DOM client-side, more easily allowing for periodically refreshing the content in a page and allowing dynamic modification of the content.  This \emph{client-side rendering}, now predominant in social media platforms such as Twitter and Instagram, is also being adopted by news outlets, such as CNN.com.  When conventional web archiving techniques, such as crawling with Heritrix, are applied to pages that render their content client-side, the JSON responses can become out of sync with the HTML page in which it is to be embedded, resulting in \emph{temporal violations} on replay.  Because the violative JSON is not directly observable in the page (i.e., in the same manner a violative embedded image is), the temporal violations can be difficult to detect.  We describe how the top level CNN.com page has used client-side rendering since April 2015 and the impact this has had on web archives. Between April 24, 2015 and July 21, 2016, we found almost 15,000 mementos with a temporal violation of more than 2 days between the base CNN.com HTML and the JSON responses used to deliver the content under the main story.  One way to mitigate this problem is to use \emph{browser-based crawling} instead of conventional crawlers like Heritrix, but browser-based crawling is currently much slower than non-browser-based tools such as Heritrix.
\end{abstract}
\maketitle

\section{Introduction}

The contents of web archives are increasingly being used as evidence -- evidence that a public figure stated support for an unpopular position \cite{Kaczynski2022Aug, AlexiMcCammond2022Aug, Roberts2022Aug}, evidence that a political candidate misrepresented their background \cite{Skalka2022Oct}, evidence that a government agency blocked public access to a report \cite{Winkie2022Aug}, and even evidence in legal cases \cite{McCarthy2018Sep, ia-legal}, among many other examples\footnote{\url{https://web.archive.org/web/20221118192715/https://archive.org/about/news-stories/search?mentions-search=Wayback+Machine}} \cite{frew-blog2022, frew-github-dataset}. For this confidence in web archives to be justified, users must be able to trust that what is replayed from a web archive is what was captured at the time or, at least, that large differences in what is displayed and what was captured are easily discernible by the user. Although we know that replay of archived webpages, or \emph{mementos}, can exhibit some differences  over time \cite{aturban-phd} due to JavaScript execution, replay software upgrades, or even archive relocation \cite{aturban-tpdl21}, we still expect that the main content of a memento reflects the content of the live webpage that was captured.  However, we have identified cases where webpages built using client-side techniques that use JSON files to populate the content of the page can produce serious temporal violations and cause web archives to reconstruct webpages containing content that did not exist at the original capture datetime.  Webpages built using these techniques can cause serious challenges for web archive crawlers and playback systems. 

We focus on the main page at CNN.com because it is a popular webpage, changes frequently, and is heavily archived. There have been previous studies, such as Atkins et al. \cite{atkins-ipres18}, that analyze what different news sites cover on their main webpages. At the time of their study, Atkins et al. had to leave CNN.com out of their analysis because mementos of the CNN.com main page were not replayable in Wayback-style replay systems \cite{berlin-blog2017}. Although mementos of the main page of CNN.com are now replayable, we have identified other issues with the playback of this webpage in certain instances. Many of these issues are not isolated to CNN.com and may be found on other webpages that use client-side webpage construction. We note that this particular example is US-centric as clients and archives outside of the US are typically redirected to the international version of CNN.com (\url{https://edition.cnn.com}). That version of the CNN main page exhibits similar issues to those that we explore here, but we focus on the \url{https://www.cnn.com} version because there are many more archived copies available for us to study. However, the issues that we uncover are not US-centric and could impact any webpage that uses similar content loading techniques.

\begin{figure*}[ht]
\centering
\begin{subfigure}{0.19\textwidth}
  \centering
  \includegraphics[width=0.99\linewidth]{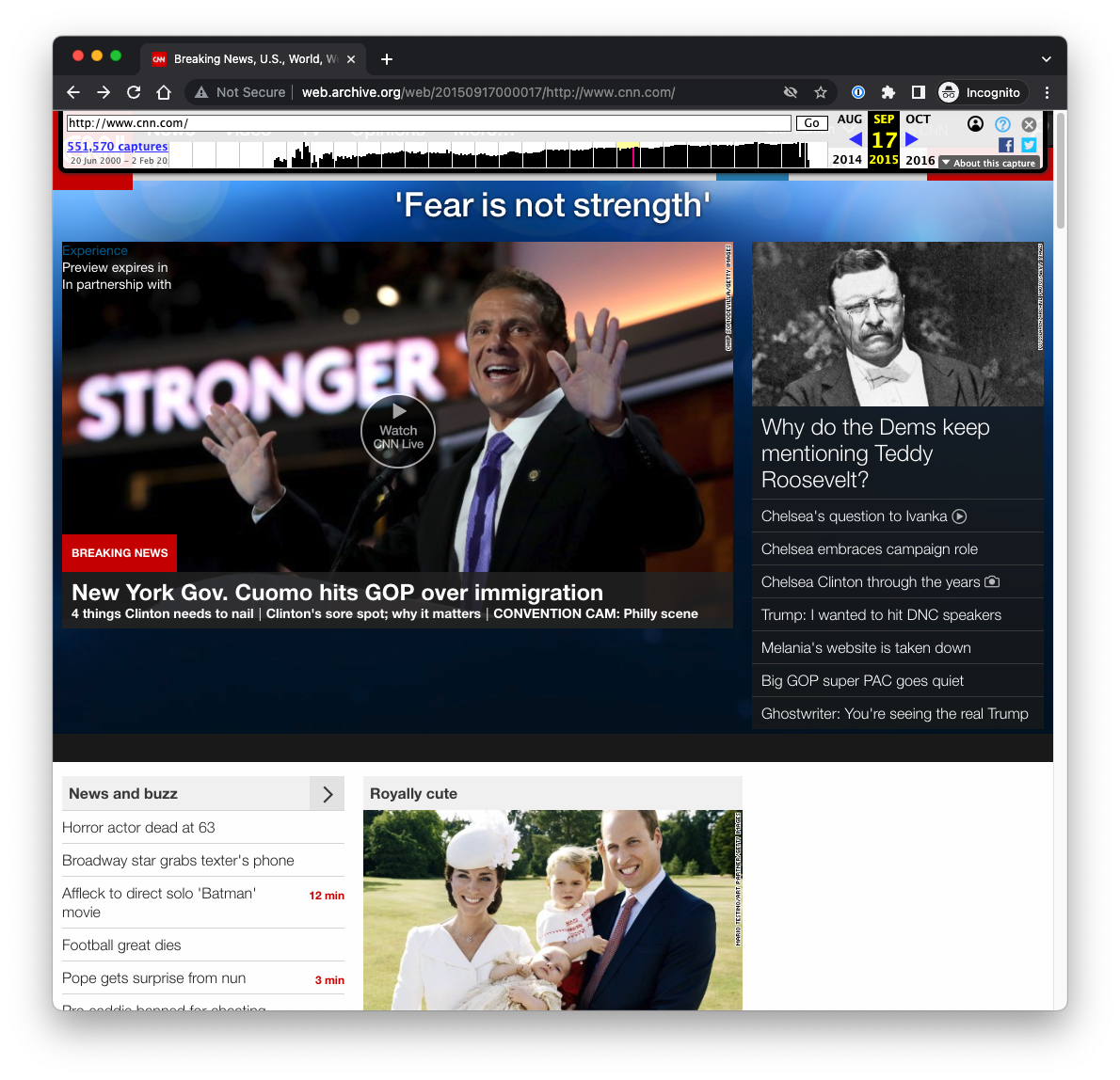}\\
  (\urim{20150917000017})
  \includegraphics[width=0.99\linewidth]{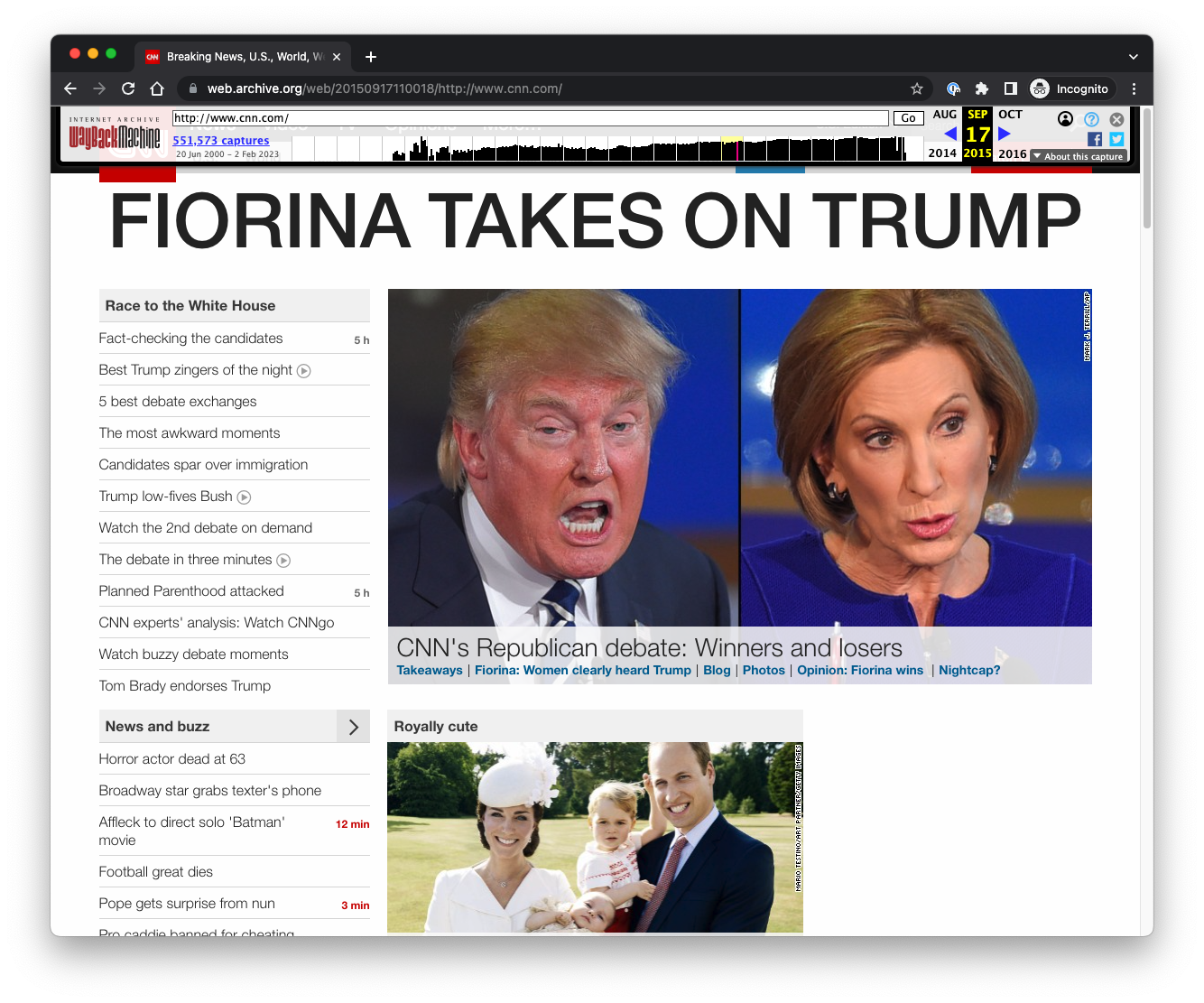}\\
  (\urim{20150917110018})
  \caption{September 17, 2015}
  \label{fig:cnn-20150917}
\end{subfigure}%
\begin{subfigure}{0.19\textwidth}
  \centering
  \includegraphics[width=0.99\linewidth]{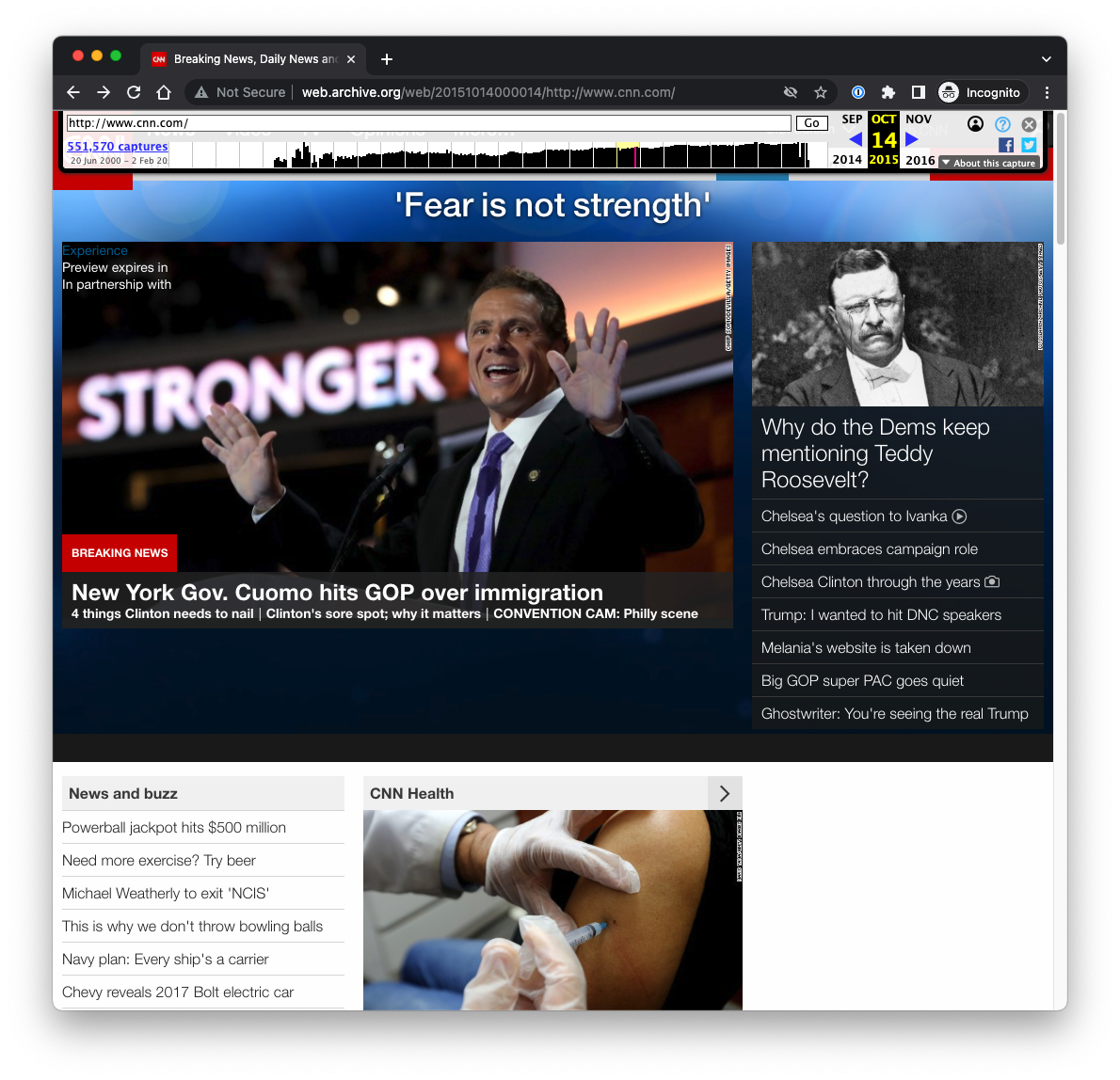}\\
  (\urim{20151014000014})
  \includegraphics[width=0.99\linewidth]{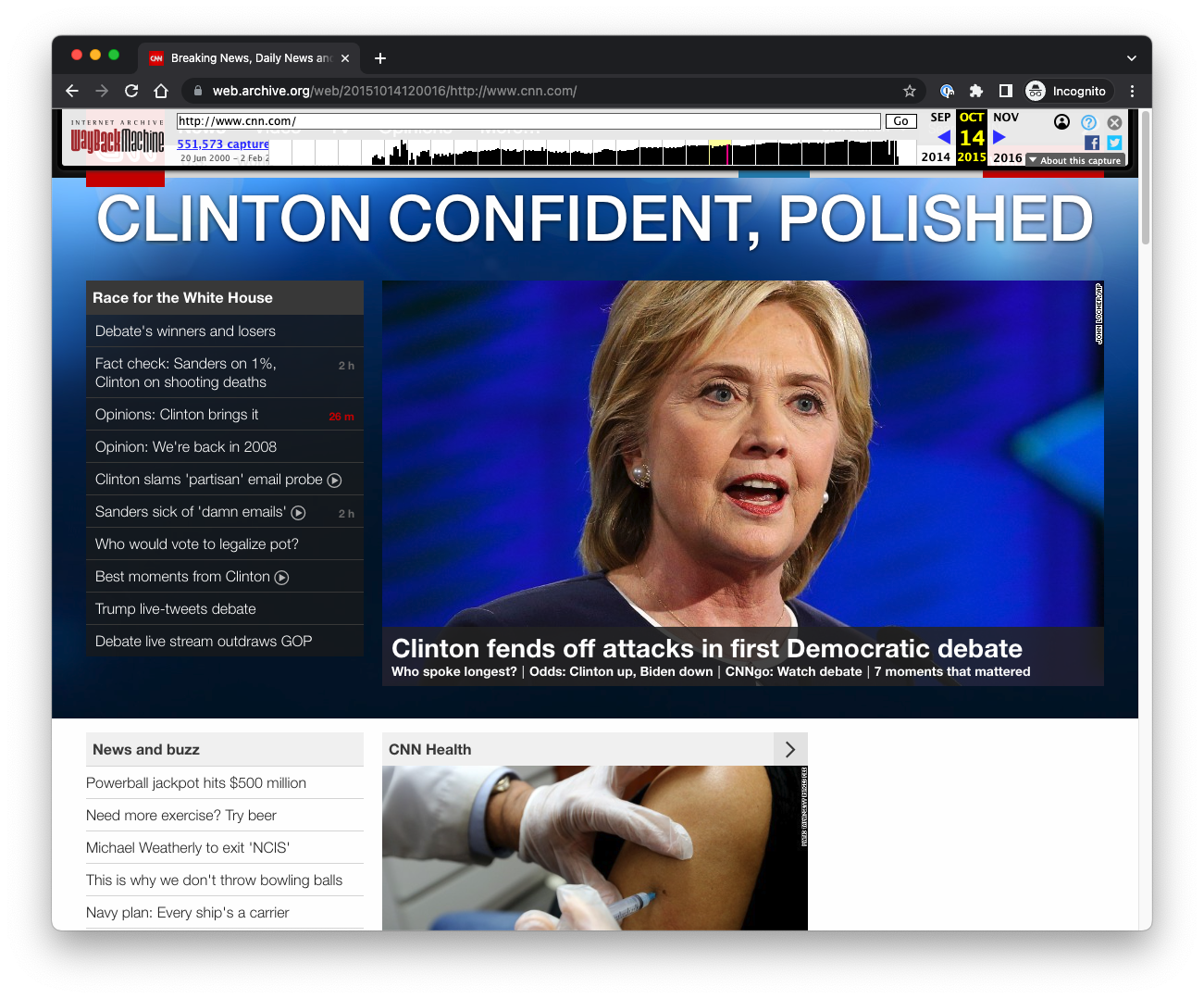}\\
  (\urim{20151014120016})
  \caption{October 14, 2015}
  \label{fig:cnn-20151014}
\end{subfigure}
\begin{subfigure}{0.19\textwidth}
  \centering
  \includegraphics[width=0.99\linewidth]{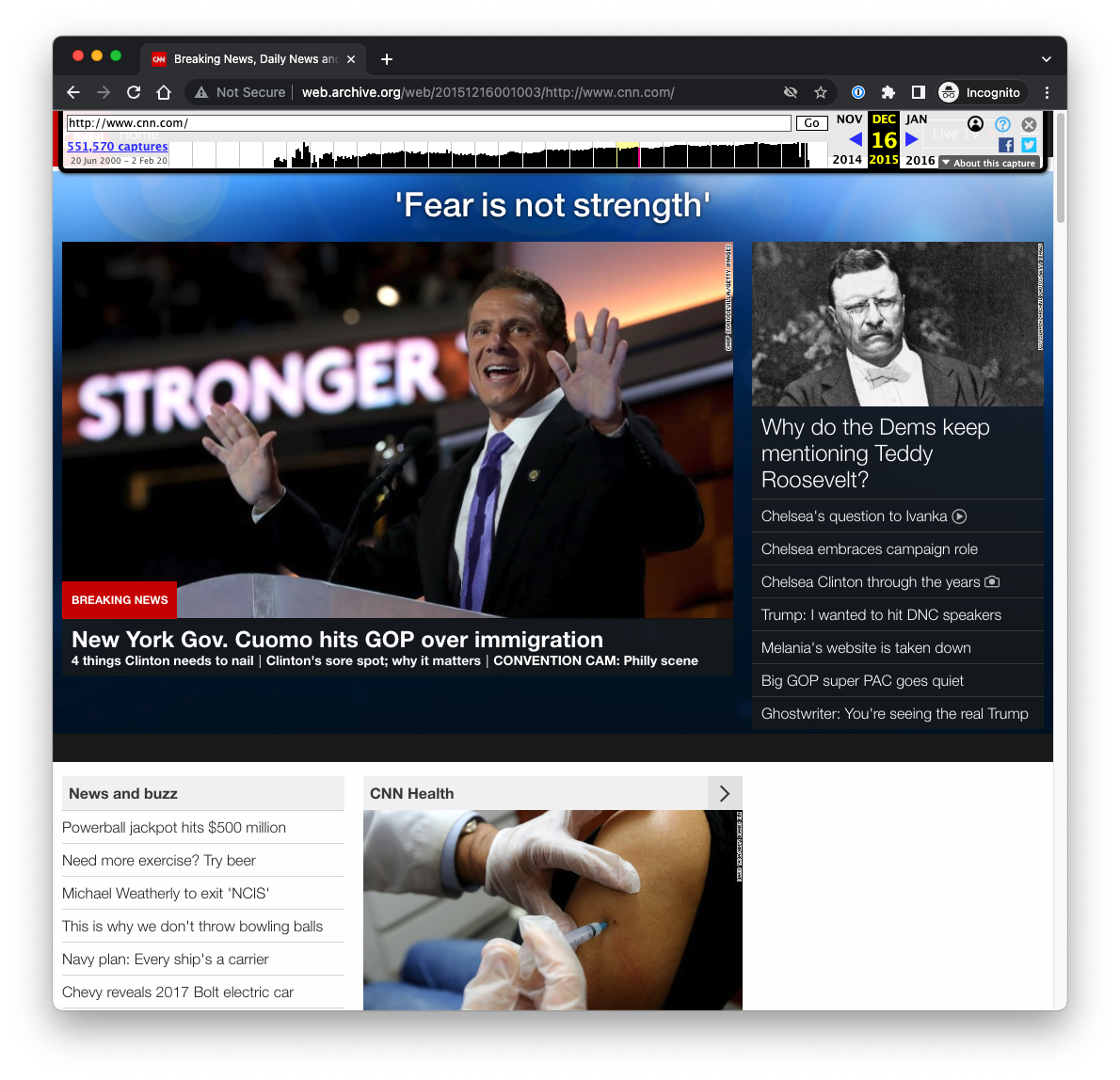}\\
  (\urim{20151216001003})
  \includegraphics[width=0.99\linewidth]{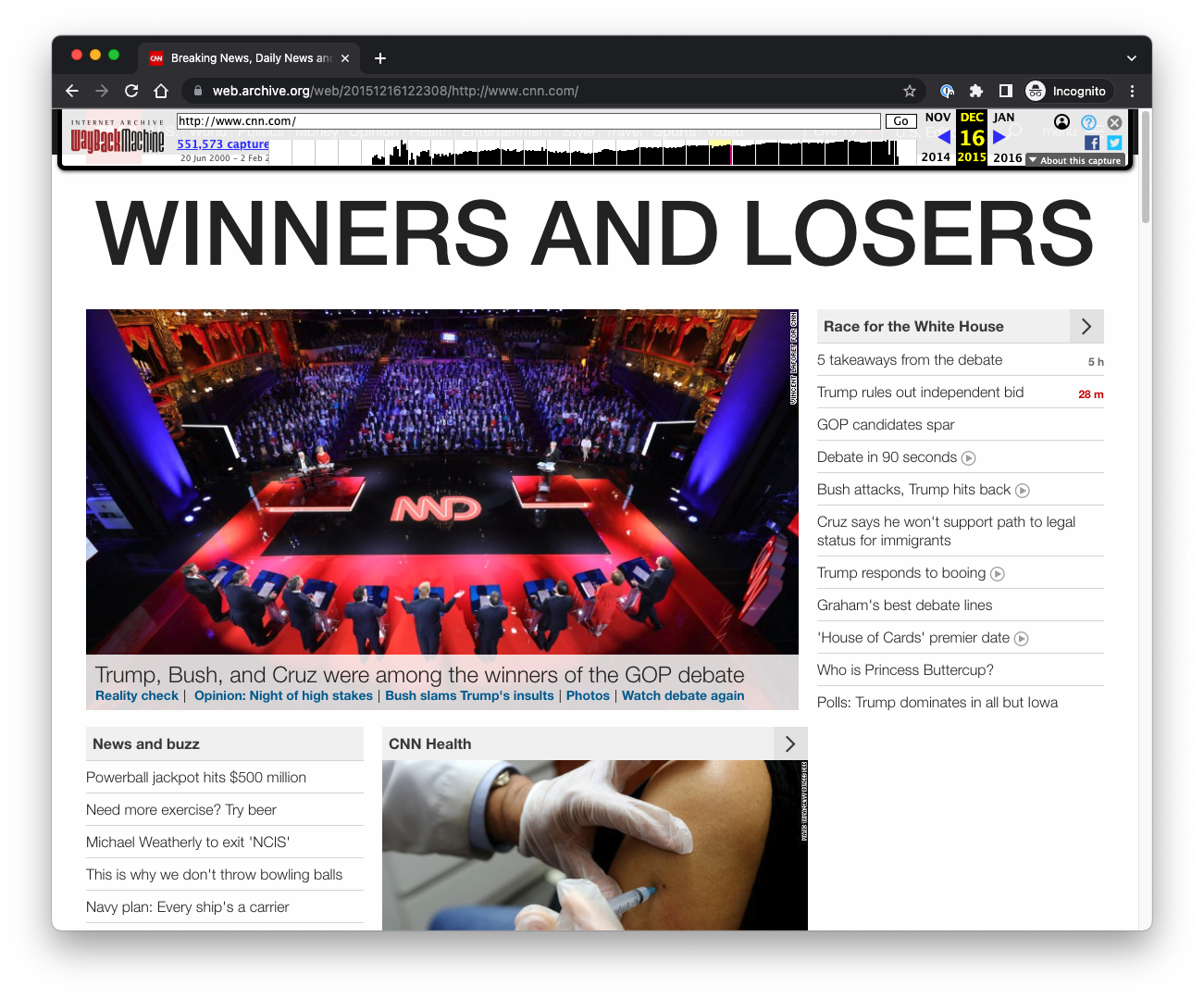}\\
  (\urim{20151216122308})
  \caption{December 16, 2015}
  \label{fig:cnn-20151216}
\end{subfigure}%
\begin{subfigure}{0.19\textwidth}
  \centering
  \includegraphics[width=0.99\linewidth]{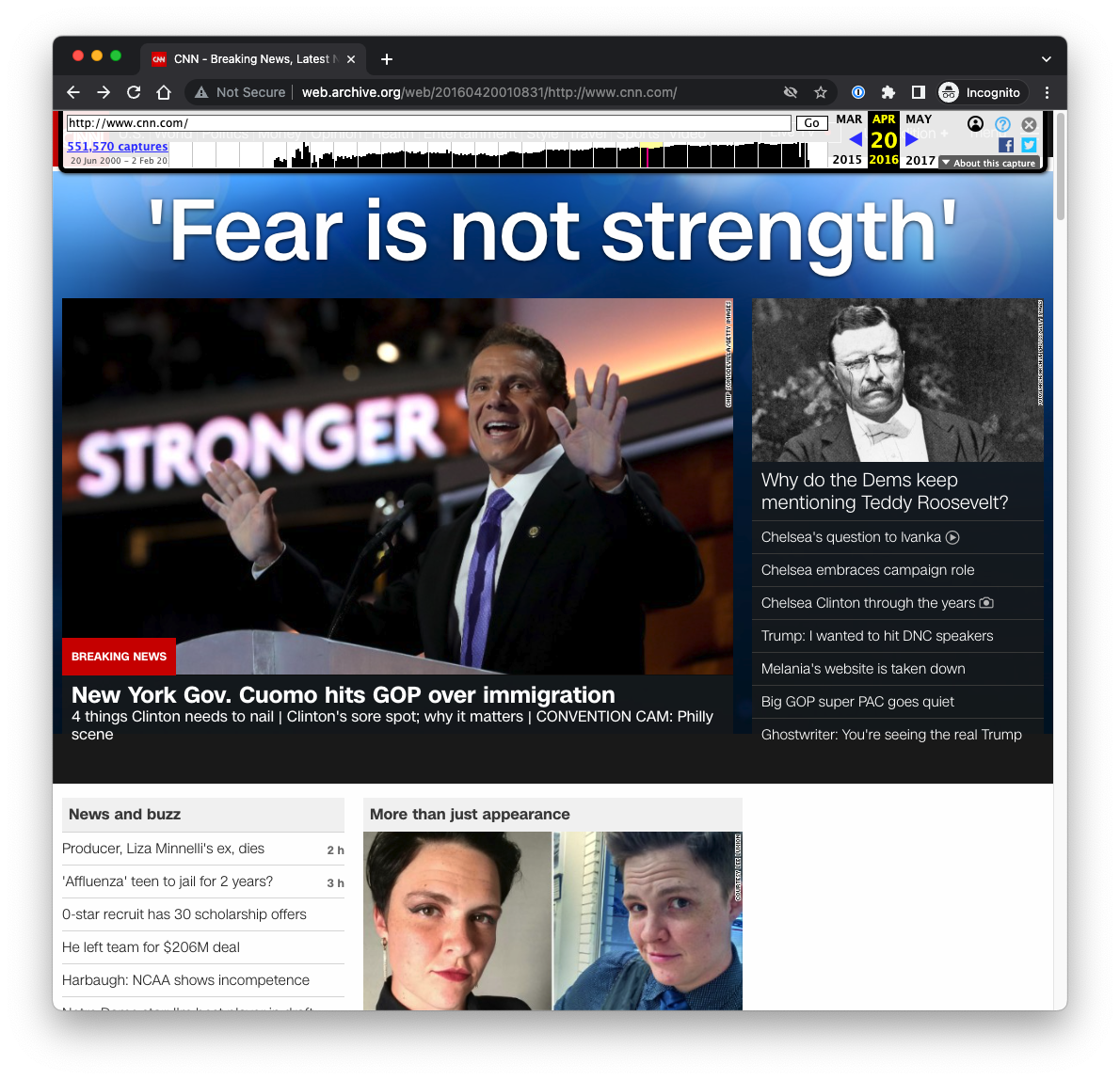}\\
  (\urim{20160420010831})
  \includegraphics[width=0.99\linewidth]{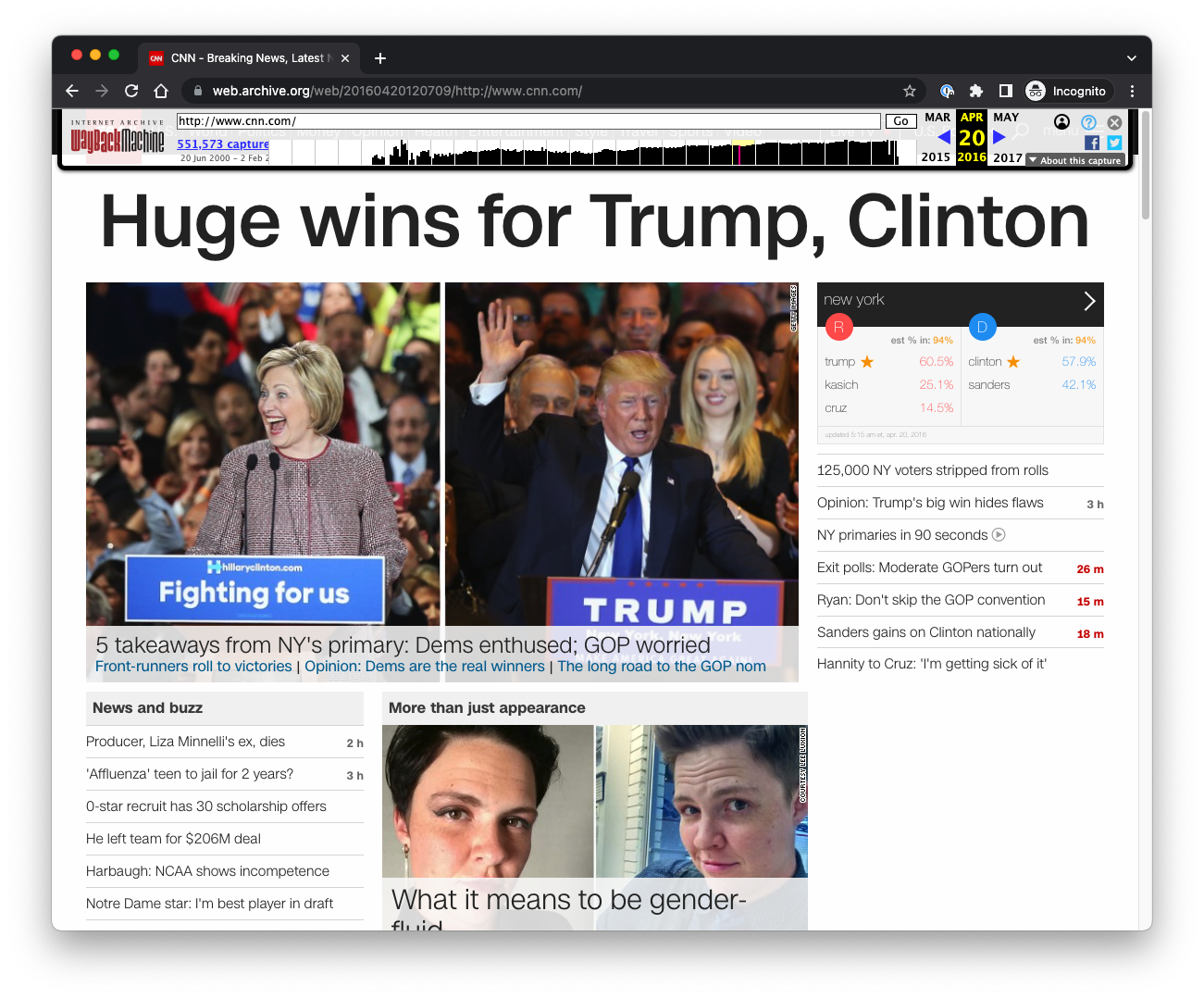}\\
  (\urim{20160420120709})
  \caption{April 20, 2016}
  \label{fig:cnn-20160420}
\end{subfigure}%
\begin{subfigure}{0.19\textwidth}
  \centering
  \includegraphics[width=0.99\linewidth]{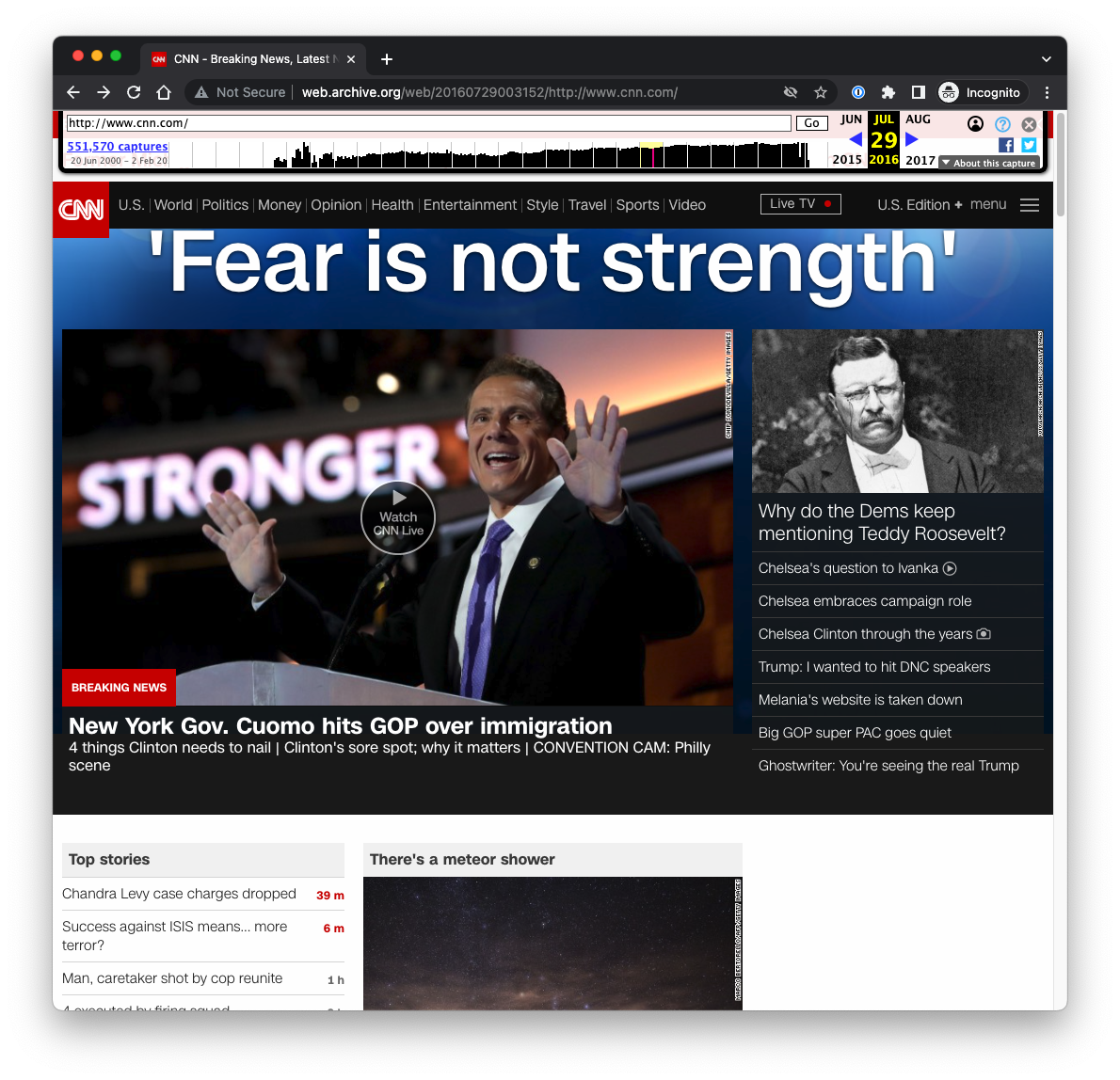}\\
  (\urim{20160729003156})
  \includegraphics[width=0.99\linewidth]{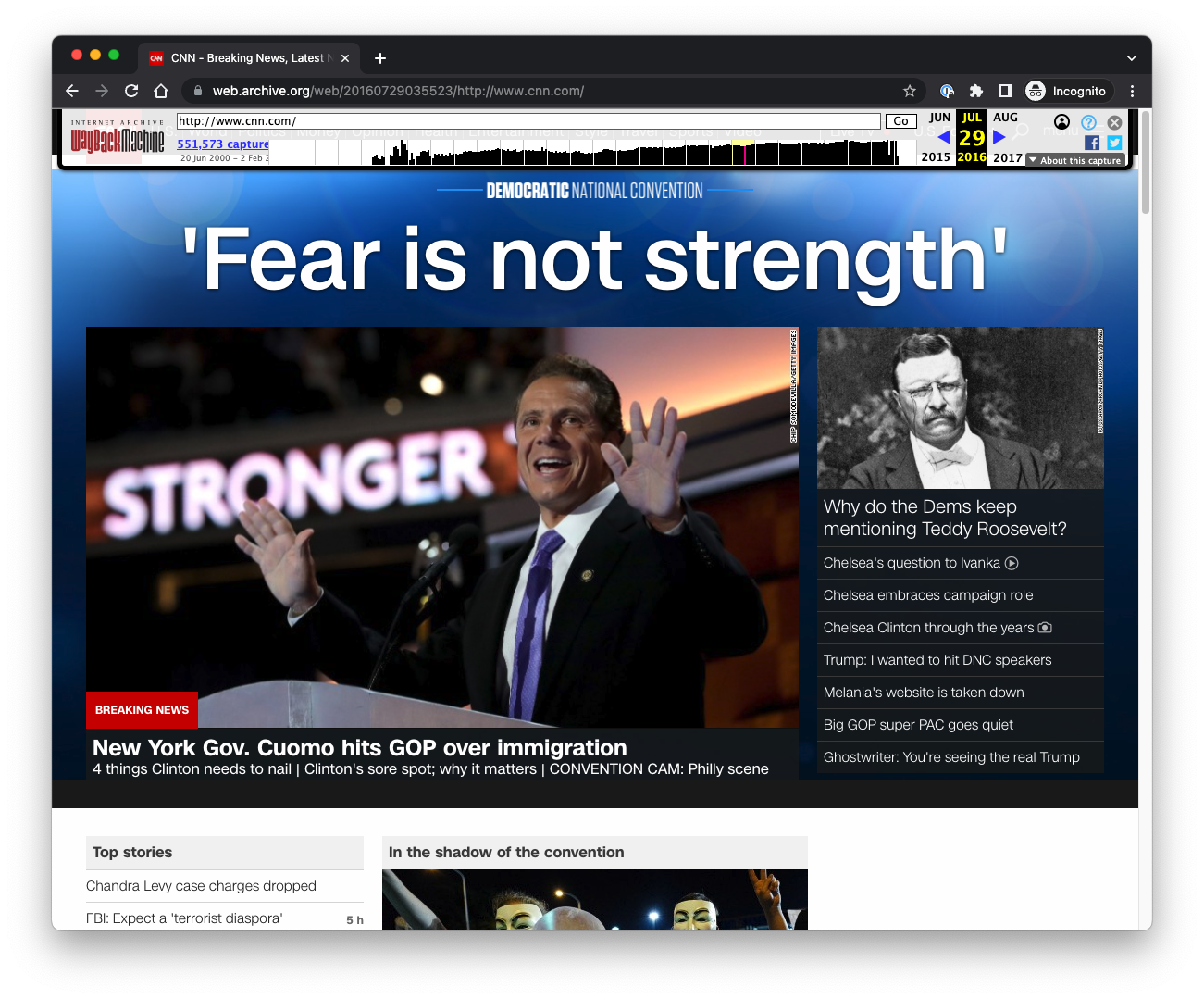}\\
  (\urim{20160729035523})
  \caption{July 29, 2016}
  \label{fig:cnn-20160729}
\end{subfigure}
\caption{Archived CNN.com webpages between September 2015 and July 2016. Top row:  archived CNN.com webpages that replay in the Wayback Machine with the same top-level content even though they were captured months apart. Bottom row: mementos from the same dates but with correct Hero stories. However, the second-level content is still temporally incorrect.}
\label{fig:cnn-hero-temporal}
\end{figure*}
Figure \ref{fig:cnn-hero-temporal} shows a sample of  screenshots of replayed mementos of CNN.com in the Internet Archive's Wayback Machine between September 2015 and July 2016.\footnote{For referencing Wayback Machine mementos in figure captions, we will include a link to the memento, but only display the datetime, which can be inserted into the general URL format \texttt{https://web.archive.org/web/\textbf{datetime}/https://www.cnn.com/}.}
The top-level content in the mementos, including the ``Hero'' (most prominent) story and the ``Top stories'' panel, is identical although the main pages were captured over a span of almost 10 months. Unfortunately, since CNN.com does not prominently display the date within the main page, these temporal violations are hard to detect. For instance, one of the columns (Figure \ref{fig:cnn-20160729}) contains the correct content for that day, but without further investigation (or knowledge of when the Democratic National Convention was held), it is difficult to tell which. Not all mementos from September 17, 2015 to July 29, 2016 display the same ``Fear is not strength'' headline, but we have found 45 that do, out of a sample of 600 mementos during that time range. 

This behavior is caused by CNN.com building out the content of the page through client-side API-driven calls for additional resources that return either HTML files or JSON files containing HTML. Further complicating the examples shown in Figure \ref{fig:cnn-hero-temporal} is the fact that during 2015-2016, the content delivery method of CNN.com was not consistent, as will be explained. These additional resources were not all archived near the same time as the base CNN.com HTML page. Since Wayback-style archives patch missing resources with those from the nearest available datetime, this can cause temporal violations upon replay. This is similar to behavior that has previously been identified with some archived Twitter pages \cite{garg-jcdl21}, where the content of the various panels on Twitter account pages are populated by JSON files that may have been archived at different times.

In this paper, we describe how the CNN.com main page is constructed and identify the JSON files that actually contain the main content. We also provide a timeline for these changes so that those studying archived CNN.com can be aware of potential issues during certain time periods. Those archiving CNN.com should ensure that these additional resources are also archived close to the datetime of the base page, most likely by using a browser-based crawler.  We also investigate the extent of temporal violations in CNN.com main pages in the Internet Archive and Archive-It. We found that almost 15,000 mementos in the Internet Archive between April 24, 2015 and July 21, 2016 have a temporal violation of over 2 days for the second level content on the CNN.com main page. This temporal violation peaked on October 8, 2015 with a difference of 90 days. Finally, we offer suggestions for how these temporal violations can be mitigated.

\section{Background and Related Work}\label{sec:background}

Web archives crawl the web in much the same way that search engines do, but instead of replacing a prior version of a crawled resource with the newest version, the web archive maintains and provides access to all versions, indexed by the time of their archive (known as the \textit{Memento-Datetime} \cite{nelson2011memento}).  Conventional web crawling and archiving, as done by applications such as Heritrix \cite{mohr:heritrix} and wget \cite{wget}, downloads an HTML page, extracts the URLs in the page (both links and embedded resources), adds those URLs to a \textit{crawl frontier}, and then crawls the next URL in the frontier.  The frontier can be indexed as \textit{breadth-first} or \textit{depth-first}, with most crawlers configured to breadth-first crawling in order to improve site-level consistency \cite{spaniol2009catch, denev2009sharc, ben2011archiving, ben2011coherence}.  For example, all of the pages linked from the top-level page (level 0) are crawled, then all the pages linked from those pages (level 1) are crawled, etc. 

This breadth-first strategy can mean that embedded resources, such as JavaScript, CSS, images, and iframes, are crawled at very different times than the page that embeds them.  These embedded resources (e.g., logos, stylesheets) are often shared between pages on a site, so crawl order is less important for them when trying to minimize the temporal spread between crawling the HTML pages for an entire site.

Recently, there has been a move to use browser-based crawling \cite{brozzler-ait, browsertrix-cloud, browsertrix-cloud-iipc} for web archiving in order to execute all the JavaScript and ensure that all the dynamically loaded resources are present.  While this approach is more complete, since all the resources needed to render the page will be archived, it is a much slower process \cite{ipres-2015:two-tier} than, for example, Heritrix, the standard crawler at the Internet Archive. 
Some services, such as archive.today \cite{archive.today} and Perma.cc \cite{zittrain2014perma}, employ browser-based crawling to archive a single page at a time and, with a single request, do not crawl multiple pages, much less an entire site.  
Since 2020, the Internet Archive's subscription service, Archive-It, has offered subscribers the use of both Heritrix and the browser-based crawler Brozzler \cite{brozzler-ait}.
The ``Save Page Now'' feature of the Internet Archive's Wayback Machine uses browser-based crawling for the single page requested \cite{Graham2019Oct}, even when the majority of the other crawls are still performed with the much faster Heritrix.  Because of the single page scope of many of these services, they have the advantage of completeness and limited temporal spread between the embedded resources (e.g., images, stylesheets) and the embedding HTML page.

Subscription-based archives, like Perma.cc and Archive-It, allow subscribers to have finer grained control of the web pages in their collections.  For example, if two different subscribers crawl the same page at the same time, there are two different copies instead of a shared single copy.  This allows subscribers to mark some pages ``private'', delete pages, and generally insulate their copy from copies in different collections elsewhere in the archive.  While pages crawled by Archive-It also become part of the larger Wayback Machine \cite{ait-faq} (i.e., with different URLs but with the same Memento-Datetime), within Archive-It, resources are not shared inter-collection.  Since users of subscription-based archives operate on a data budget, their collections can be sparse (i.e., very few copies) even for otherwise popular pages, like twitter.com or CNN.com.  Without inter-collection sharing, replaying a page in Archive-It will be limited to the resources within that specific collection.

The lack of inter-collection sharing combined with the standard procedure of how Wayback Machines and their work-alikes (e.g., Open Wayback \cite{openwayback:github} and PyWB \cite{pywb}) resolve temporal preferences can lead to a large \textit{temporal spread} \cite{ainsworth:composite:tr}.  Links and embedded resources in a replayed archived web page are rewritten to specify the same Memento-Datetime as the embedding page.  In practice, those resources are rarely archived at the same Memento-Datetime, so the Wayback Machine issues an HTTP redirect to the temporally closest archived version, whether it is in the future or the past \cite{jcdl13:drift, ijdl:jcdl13}.  This spread can result in a \textit{temporal violation}, a combination of embedding page and embedded resources that never existed together on the live web \cite{ht15:as-presented}.  By itself, a large temporal spread is not necessarily a temporal violation (e.g., an image logo or stylesheet that has not changed for 10 years will still be valid even if the temporal spread is 9 years), but in practice, the greater the temporal spread, the more likely it is there is a temporal violation.

A recent web page design trend has significant impacts on web archiving and successfully replaying archived web pages.  Increasingly, dereferencing a URL returns not a complete HTML page with all of the information assembled server-side, but instead the server returns a skeleton page of HTML and JavaScript, which then issues a series of API calls 
to other web servers.  These API responses return data (often JSON), which is then used by the JavaScript to build out the rest of the page, a process known as \emph{client-side rendering} (CSR) \cite{csr}.  With a non-browser-based crawler such as Heritrix, these API calls are not made and their responses are not archived.  This leads to incomplete and erroneous archived pages when replayed, such as the described problems with Twitter \cite{garg-jcdl21} and Instagram \cite{instagram:archived, bragg-jcdl23}.  This client-side rendering of pages is well-suited for social media, with their highly interactive and constantly updating UIs.  However, we have discovered client-side rendering has also been adopted by news web sites, such as CNN.com, and there are significant implications for replaying their archived pages. 

\section{Client-Side Construction of CNN.com}
\label{sec:construction}

The Internet Archive's Wayback Machine contains almost 200,000 mementos with HTTP status code 200 for \texttt{www.cnn.com} since 2010.  Using this valuable resource, we analyzed several years' worth of mementos from CNN.com and found that the client-side construction of the main page began on April 24, 2015.\footnote{The last memento before this construction is \urim{20150424134720}, and the first memento using this new construction is \urim{20150424150304}.} 
We identified several \texttt{zone-manager.html} files being requested via CSR that contained the HTML content used to build the various sections, or zones, on the CNN main page.\footnote{The earliest zone-manager file found in the Internet Archive is \url{https://web.archive.org/web/20150427192038/http://www.cnn.com/data/ocs/section/index.html:homepage4-zone-5/views/zones/common/zone-manager.html}.}  The URL of these files is in the format\\
\texttt{http://www.cnn.com/data/ocs/section/\textbf{\{baseUri,uri\}:id}/views/zones/common/zone-manager.html}, with \\
\texttt{baseUri}, \texttt{uri}, and \texttt{id} defined in the \texttt{CNN.Zones} variable (see Figure \ref{fig:lst:zones}).\footnote{For example, zone \textsf{homepage-injection-zone-1} is defined in \url{http://www.cnn.com/data/ocs/section/_homepage-zone-injection/index.html:homepage-injection-zone-1/views/zones/common/zone-manager.html}, and zone \textsf{homepage3-zone-1} is defined in \url{http://www.cnn.com/data/ocs/section/index.html:homepage3-zone-1/views/zones/common/zone-manager.html}.}  The specific files that are loaded is based on the minimum width of the browser window when the JavaScript code is executed. Our analysis is based on the desktop version of CNN.com, so we are only concerned with widths of at least 800 pixels.
\begin{figure}[ht]
\begin{lstlisting}[basicstyle=\ttfamily]
CNN.Zones = {
 (*\textcolor{gray}{[...]}*)
 "zones": {
  "baseUri": "index.html",
  "minWidth": {
   "0": (*\textcolor{gray}{[...]}*),
   "640": (*\textcolor{gray}{[...]}*),
   "800": [{"id": "homepage-injection-zone-1", 
           "uri": "_homepage-zone-injection/index.html"}, 
           {"id": "homepage1-zone-1"}, 
           {"id": "homepage-injection-zone-2",
           "uri": "_homepage-zone-injection/index.html"},
           {"id": "homepage2-zone-1"}, 
           {"id": "homepage3-zone-1"}, 
           {"id": "homepage4-zone-1"}, 
           {"id": "homepage4-zone-2"},
           {"id": "homepage4-zone-3"},
           {"id": "homepage4-zone-4"}, 
           {"id": "homepage4-zone-5"}, 
           {"id": "homepage4-zone-6"},
           {"id": "homepage4-zone-7"}]
}}};
\end{lstlisting}
\caption{JavaScript code included in the base HTML page of CNN.com from June 18, 2020 (\href{https://web.archive.org/web/20200618234848/http://www.cnn.com/}{20200618234848}) defining the webpage zones, including \texttt{baseUri}, \texttt{uri}, and \texttt{id} variables}
\label{fig:lst:zones}
\end{figure}

\begin{figure*}[tp!]
  \centering
  \includegraphics[height=0.9\textheight]{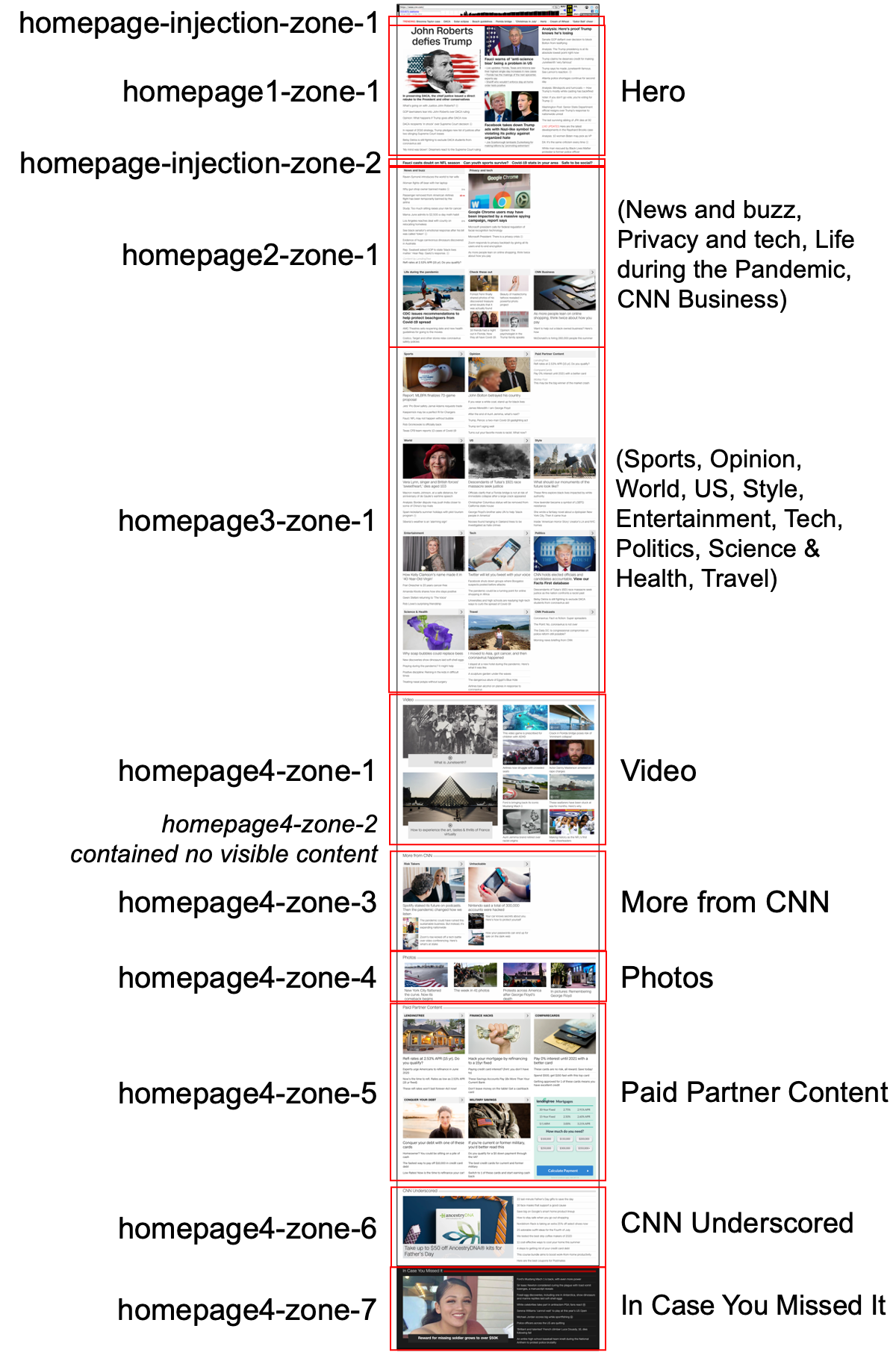}
  \caption{Archived CNN main page from June 18, 2020 (\urim{20200618234848}),  annotated with zone names and labels. Note that \textsf{homepage4-zone-2} contained no visible content.}
\label{fig:cnn-layout}
\end{figure*}

Figure \ref{fig:cnn-layout} shows a memento from June 18, 2020 annotated to illustrate this construction.  
The zone name is shown to the left of each section, and the \texttt{data-zone-label} (found in the HTML) is shown on the right.
For those zones without a \texttt{data-zone-label}, we provide a sample of the section headings.
In this memento, 11 zone-manager files are loaded, including those for \textsf{homepage-injection-zone-1} and \textsf{homepage-injection-zone-2}, which contain the HTML code for the thin separator banners. The main news content appears in zones \textsf{homepage1-zone-1}, \textsf{homepage2-zone-1}, and \textsf{homepage3-zone-1}, with the other zones loading more promotional-type content. Because we are focused on consequential temporal differences, we will discuss only these top three zones for the remainder of this paper. 

To carry out this analysis, we first used a headless browser, selenium-wire,\footnote{\url{https://pypi.org/project/selenium-wire/}} to load mementos of http://www.cnn.com/ from the Internet Archive and capture what requests were made along with their status codes. Using this method, we requested a total of 1131 mementos between 2013--2020 to study the request pattern of the main page of CNN.com over time. As mentioned earlier, we were able to pinpoint the introduction of the zone-manager files on April 24, 2015. After this point, all zones except for \textsf{homepage1-zone-1} were consistently requested via CSR. 

However, it was not until November 1, 2016 that \textsf{homepage1-zone-1} was always requested via CSR. (This was also the first time that the two homepage-injection zones appeared.) We expected to find that mementos that did not request \textsf{homepage1-zone-1} would be missing the Hero content, but upon inspection of the replay in the Wayback Machine, we found that the Hero content was temporally correct. To investigate this further, we downloaded the raw HTML of 1316 http://www.cnn.com/ mementos between 2010 and 2022, using the \texttt{id\_} option to the Wayback Machine.
We found that when \textsf{homepage1-zone-1} was \emph{not} loaded via CSR, a \texttt{<section>} tag with the attribute \texttt{id="homepage1-zone-1"} would be present in the HTML, containing the Hero content. A snippet of this is shown in Figure \ref{fig:lst:id_homepage1}. Line 1 shows the \texttt{id="homepage1-zone-1"} attribute and \texttt{data-zone-label="Hero"}. The headline ``Bombshell find at Yellowstone'' appears on line 4, formatted with \texttt{<h2>} and \texttt{<strong>} tags. When \textsf{homepage1-zone-1} was requested via CSR, the \\
\texttt{id="homepage1-zone-1"} attribute was not present in the HTML. This was verified by examining the raw HTML of 1069 mementos between April 24, 2015 and November 1, 2016 and comparing it to the requests logged by the headless browser from that same period.
\begin{figure}[ht]
\begin{lstlisting}[basicstyle=\ttfamily]
(*\bfseries \textcolor{red}{<section}*) class="zn zn-homepage1-zone-1 zn-left-fluid-right-stack zn--idx-0 t-light zn-loaded zn-left-fluid-right-stack zn-has-multiple-containers zn-has-3-containers" data-eq-pts="xsmall: 0, medium: 460, large: 780, full16x9: 1100" (*\bfseries \textcolor{red}{id="homepage1-zone-1"}*) data-vr-zone="zone-0-0" (*\bfseries \textcolor{red}{data-zone-label="Hero"}*) data-containers="3">
(*\textcolor{gray}{[...]}*)
<h2 data-analytics="_list-hierarchical-xs_article_" class="banner-text js-screaming-banner-text screaming-banner-text">
<strong>(*\bfseries \textcolor{blue}{Bombshell find at Yellowstone}*)</strong></h2></a>
\end{lstlisting}
\caption{HTML snippet that includes the \texttt{id="homepage1-zone-1"} attribute and Hero content, from April 24, 2015 (\urim{20150424150304}).}
\label{fig:lst:id_homepage1}
\end{figure}
 
The zone-manager files have undergone several formatting and file extension changes. In October 2016, the \texttt{zone-manager.html} files were changed to JSON files containing the HTML code as a value, and named\\
\texttt{zone-manager.izl.json}. Then in January 2017, the JSON formatted zone-manager files were renamed as\\
\texttt{zone-manager.izl}. An example of a \textsf{homepage1-zone-1} \texttt{zone-manager.izl} is shown in Figure \ref{fig:lst:izl-example}, showing the ``John Roberts defies Trump'' headline from Figure \ref{fig:cnn-layout} on line 5.
\begin{figure}[tb]
\begin{lstlisting}[basicstyle=\ttfamily]
{"izlData": {"cards": []},
 "html": 
   "(*\bfseries \textcolor{red}{<section}*) class=\"zn zn-homepage1-zone-1 zn-left-fluid zn--idx- zn--ordinary t-light zn-left-fluid-share zn-has-multiple-containers zn-has-3-containers\" data-eq-pts=\"xsmall: 0, medium: 460, large: 780, full16x9: 1100\" (*\bfseries \textcolor{red}{id=}*)\"(*\bfseries \textcolor{red}{homepage1-zone-1}*)\" data-vr-zone=\"zone-0-0\" (*\bfseries \textcolor{red}{data-zone-label=}*)\"(*\bfseries \textcolor{red}{Hero}*)\" data-containers=\"3\">
(*\textcolor{gray}{[...]}*)
  <h2 class=\"banner-text screaming-banner-text banner-text-size--char-26\" data-analytics=\"_list-hierarchical-xs_article_\">(*\bfseries \textcolor{blue}{John Roberts defies Trump}*)</h2></a>
\end{lstlisting}
\caption{Snippet of \href{https://web.archive.org/web/20200618234850/https://www.cnn.com/data/ocs/section/index.html:homepage1-zone-1/views/zones/common/zone-manager.izl}{\texttt{zone-manager.izl}} for \textsf{homepage1-zone-1} on June 18, 2020, showing HTML code inside a JSON formatted file, containing the headline ``John Roberts defies Trump'', seen in Figure \ref{fig:cnn-layout}.}
\label{fig:lst:izl-example}
\end{figure}
Table \ref{tab:timeline} summarizes the timeline of the filename and format changes.
\begin{table}
  \caption{CNN.com Zone Manager Timeline}
  \label{tab:timeline}
  \begin{tabular}{ll}
    \toprule
    Date&Zone Content Delivery\\
    \midrule
    Feb 18, 2015 & content in base HTML divided into zones\\
    Apr 24, 2015 & content other than Hero zone requested via CSR in \texttt{zone-manager.html} files\\
    Sep 17, 2015 & Hero zone (\textsf{homepage1-zone-1}) \emph{sometimes} requested via CSR\\
    Oct 18, 2016 & zone-manager format changed to \texttt{.izl.json}\\
    Nov 1, 2016 & all zones requested via CSR \\
    Jan 31, 2017 & zone-manager extension changed to \texttt{.izl}\\
  \bottomrule
\end{tabular}
\end{table}

\section{Impact on Web Archives}

CNN's client-side rendering model conflicts with standard web archiving models. Since the webpage content is built out with blocks of HTML or JSON, if those zone-manager files are not archived at the same time as the base HTML, there will be temporal violations. Upon playback, the CNN.com with the datetime indicated in the archive banner will be populated with data from files that were archived at a different time. 
Further, CNN.com no longer displays the date on the main page or includes it in any of the content to be displayed in its zones, so these temporal violations will be difficult to detect.
Because of the time-sensitive nature of the content in these files, the violations are more serious the greater the distance between captures, especially with content being populated from the future. 

We used MemGator \cite{jcdl-2016:alam:memgator} in November 2022 to gather the TimeMap (list of mementos) for http://www.cnn.com/ from public web archives with Memento-Datetimes after April 24, 2015. Table \ref{tab:num_mementos} shows that the Internet Archive (web.archive.org) and Archive-It (wayback.archive-it.org) have the most mementos during this time period by far. The focus of this particular study, CNN.com, uses geolocation to determine which webpage to deliver, so web archives outside of the US (including archive.today, arquivo.pt, webarchive.org.uk, and web.archive.org.au) will likely not actually have captures of http://www.cnn.com/, but rather http://edition.cnn.com/.  This is a function of what the web server delivers to the archive rather than of the archive or its capture technology.
\begin{table}
  \caption{Number of mementos for http://www.cnn.com/ after April 24, 2015 in public web archives (via MemGator)}
  \label{tab:num_mementos}
  \begin{tabular}{rr}
    \toprule
    Archive&Mementos\\
    \midrule
    web.archive.org & 483,611\\
    wayback.archive-it.org & 35,941\\
    archive.today & 3276\\
    webarchive.loc.gov & 1666\\
    arquivo.pt & 1439\\
    swap.stanford.edu & 495\\
    perma.cc & 139\\
    www.webarchive.org.uk & 11\\
    web.archive.org.au & 8\\
  \bottomrule
\end{tabular}
\end{table}

\begin{figure}[t]
\begin{subfigure}{0.47\linewidth}
\centering
  \includegraphics[width=0.99\linewidth]{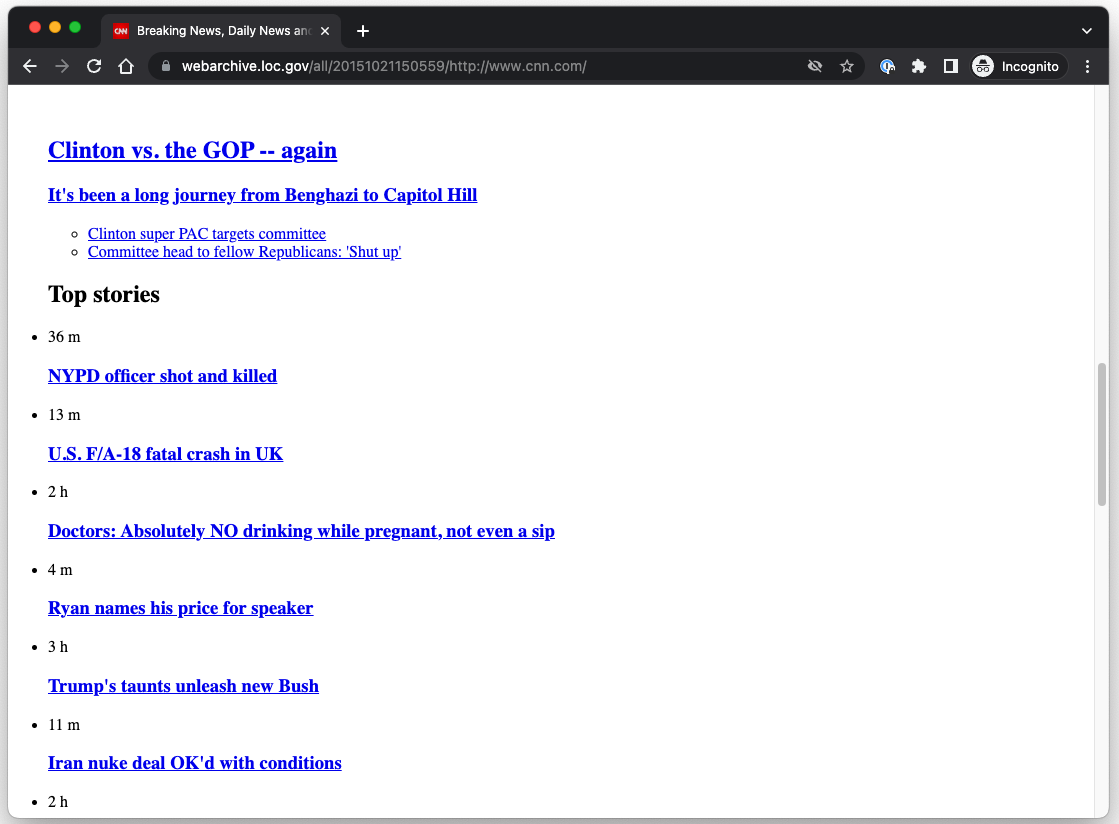}
  \caption{CNN.com archived with a conventional crawler} 
\label{fig:loc}
\end{subfigure}
\hspace{6pt}
\begin{subfigure}{0.47\linewidth}
\centering
  \includegraphics[width=0.99\linewidth]{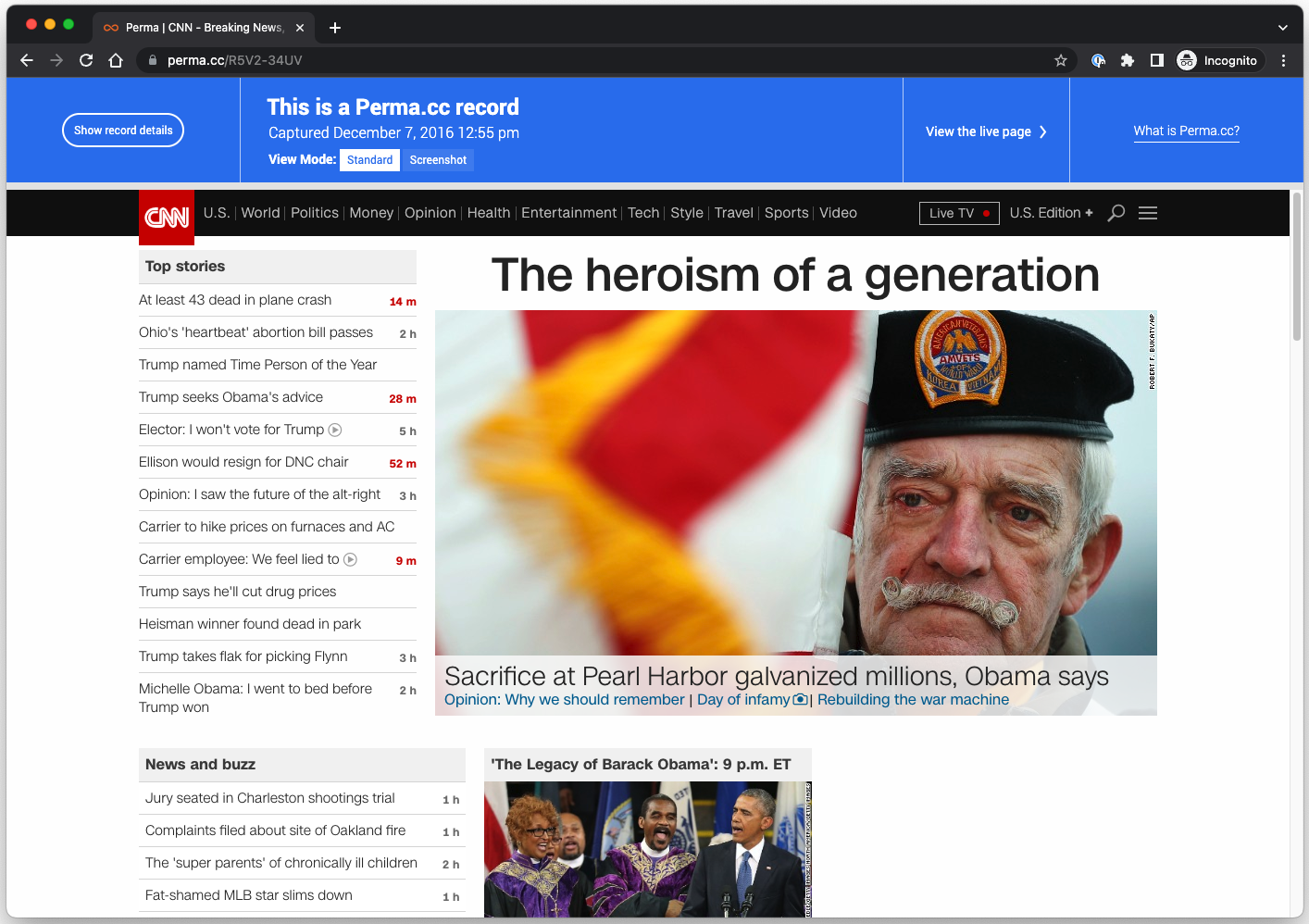}
  \caption{CNN.com archived with a browser-based crawler} 
\label{fig:perma}
\end{subfigure}
\caption{CNN.com replayed from (a) the Library of Congress web archive (webarchive.loc.gov), archived on \href{https://webarchive.loc.gov/all/20161129193435/http://www.cnn.com/}{Oct 21, 2015} using a conventional crawler, and (b) Perma.cc, archived on \href{https://perma.cc/R5V2-34UV}{Dec 7, 2016} using a browser-based crawler.} \label{fig:other-archives}
\end{figure}

We also used MemGator to look for mementos of zone-manager files in archives other than Internet Archive or Archive-It but found only a few. The Library of Congress web archive (webarchive.loc.gov) has 11 zone-manager.izl files, all from June 22, 2019. The only other zone-manager memento was for a homepage4-zone-1 memento from March 15, 2016 in Arquivo.pt. Because these zone-manager files hold the actual content from the CNN.com main page, without these files, mementos of CNN.com in other archives after April 24, 2015 will either be damaged and show little or no content or will have serious temporal violations. For instance, Figure \ref{fig:loc} shows the replay of a CNN.com memento from the Library of Congress web archive from October 12, 2015. 
The image is scrolled down to show the display of the Hero and Top stories sections that are included in the HTML, but the CSS and other zone content is not loaded.
Web archives that used browser-based crawling, such as archive.today, Perma.cc (example from December 7, 2016 in Figure \ref{fig:perma}), and the Internet Archive's Save Page Now feature, will not have the same problems that non-browser-based crawlers have, because the browser will request and archive all of the zone files when the DOM is rendered for that page.  
As stated before, the drawback to browser-based crawling is that it is much slower than conventional crawlers, such as Heritrix, and is not suitable for ``wide'' crawls of entire sites.  

Because they have the most mementos, we focus the rest of our analysis on CNN.com mementos replayed from the Internet Archive and from particular collections in Archive-It. 

\subsection{Internet Archive}

\begin{figure}[ht]
  \centering
  \includegraphics[width=\linewidth,trim=0 0 70 0, clip]{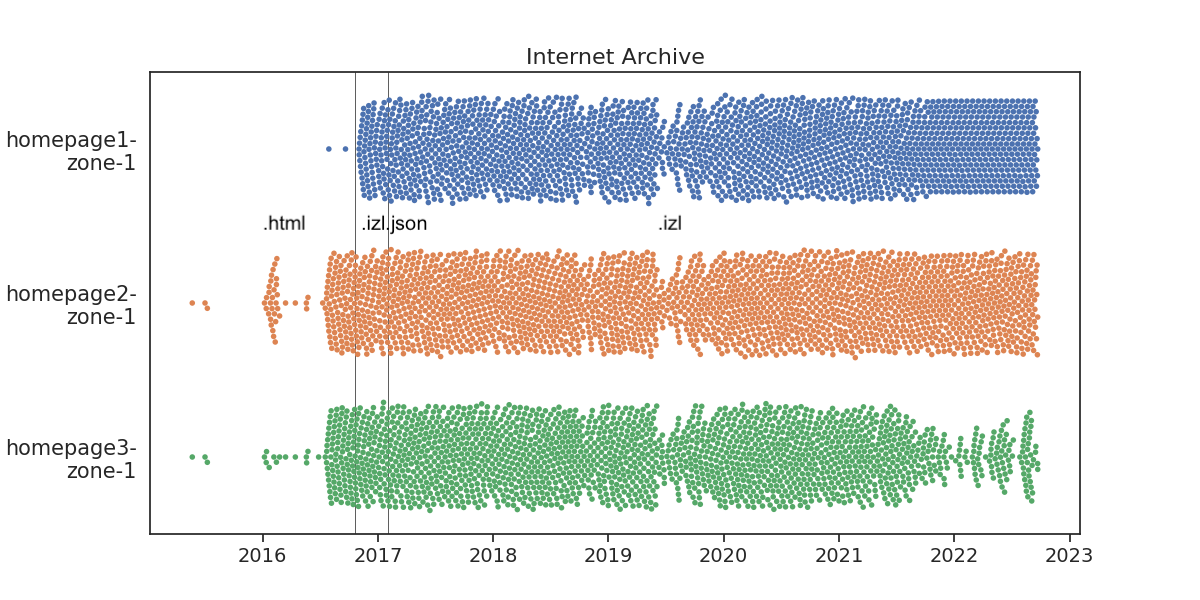}
  \caption{Mementos in the Internet Archive for homepage1-zone-1, homepage2-zone-1, homepage3-zone-1 zone-manager files. Each dot represents at least one memento on that day.}
\label{fig:alldates}
\end{figure}

Based on data from the Internet Archive's CDX API, there are 172,776 mementos with an HTTP 200 status code for http://www.cnn.com/ in the Internet Archive between April 24, 2015 and September 23, 2022. Figure \ref{fig:alldates} shows a plot of the mementos available in IA for the \textsf{homepage1-zone-1}, \textsf{homepage2-zone-1}, and \textsf{homepage3-zone-1} zone-manager files. The Memento-Datetime is along the x-axis and the zone files are separated along the y-axis. Each dot represents the presence of at least one memento on that day. The stacking of dots of the same color has no meaning, but is just to prevent over-plotting. We do not show the www.cnn.com mementos in this plot because there is at least one memento every day. The figure is annotated with the filename extensions used at different time periods.

\begin{figure}[ht]
\begin{subfigure}{0.99\linewidth}
  \centering
  \includegraphics[width=\linewidth]{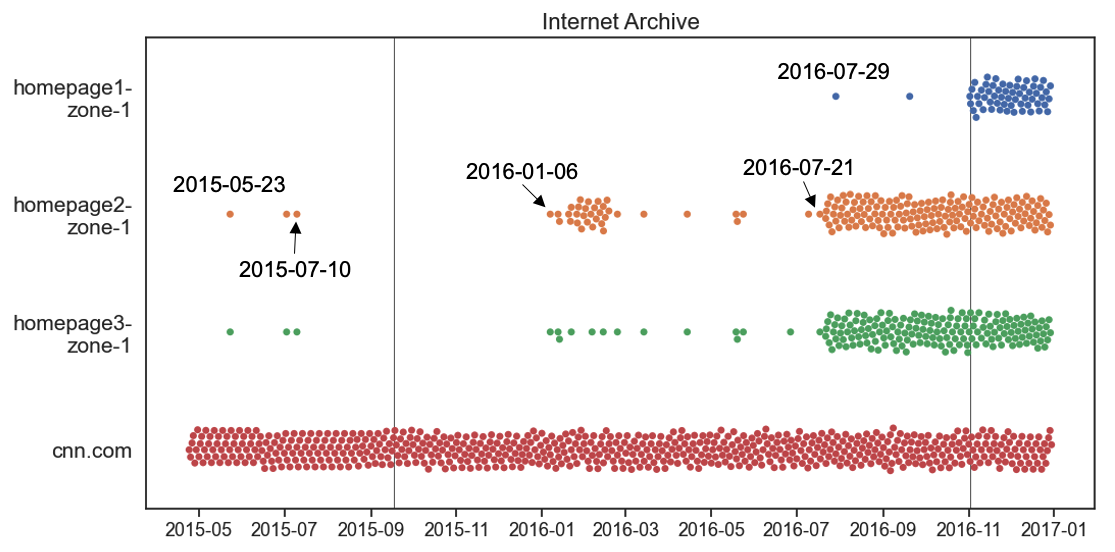}
  \caption{Mementos in the Internet Archive for CNN.com and top three zones.}
    \label{fig:IA-urims-hp123}
\end{subfigure}
\begin{subfigure}{0.99\linewidth}
  \centering
  \includegraphics[width=\linewidth,trim=0 20 70 0, clip]{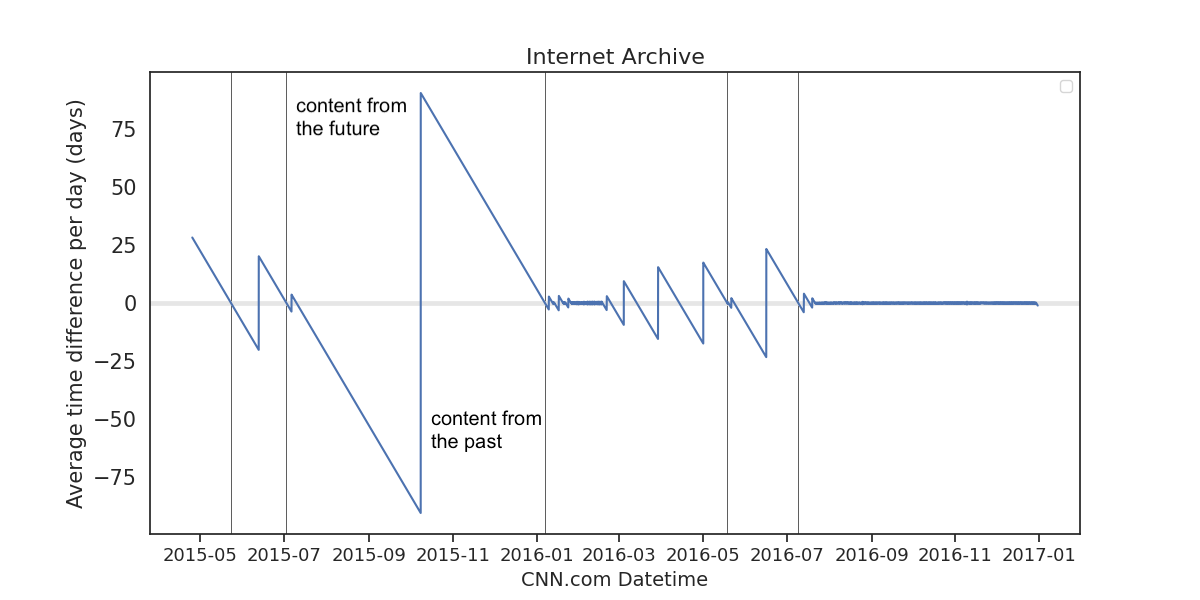}
  \caption{Average difference in days between the CNN.com datetime and the nearest memento of the \textsf{homepage2-zone-1} zone-manager file in the Internet Archive. }
\label{fig:IA-hp2-diff}
\end{subfigure}
\caption{Analysis of Internet Archive Mementos between April 24, 2015 and December 31, 2016. (a) One dot for each day with a memento, (b) Difference between www.cnn.com and nearest \textsf{homepage2-zone-1} memento.}
\label{fig:IA-201504-201612}
\end{figure}

From Figure \ref{fig:alldates}, we can see that after 2017, there is fairly consistent presence of the zone-manager mementos, so we focus our analysis on the time period between April 24, 2015 and December 31, 2016.  Figure \ref{fig:IA-201504-201612} shows two different views of this time period. 
Figure \ref{fig:IA-urims-hp123} shows the mementos available in the Internet Archive for www.cnn.com and the top three zones, with several dates annotated. The first vertical line is the first date where we see \textsf{homepage1-zone-1} requested (September 17, 2015), and the second vertical line is when we see all zones regularly loaded via CSR (November 1, 2016). As noted in Section \ref{sec:construction}, \textsf{homepage1-zone-1} was only occasionally requested via CSR during this time period. For now, we concentrate our analysis on the two main zones that were always requested via CSR after April 2015 and will return to \textsf{homepage1-zone-1} later. 
For the other two zones, we know that during this time period these zones were always loaded via CSR, so any white space on those rows in Figure \ref{fig:IA-urims-hp123} indicates a temporal violation. This means that when a memento for CNN.com is replayed, there will be no nearby memento for the zone, so the content will come from the nearest zone memento, which could be days (or even months) in the past or future. To simplify the discussion, in the next analysis we focus on \textsf{homepage2-zone-1}, as the results will be similar to that of \textsf{homepage3-zone-1}. 

In Figure \ref{fig:IA-hp2-diff}, we show the average difference (in days) per day between the requested CNN.com datetime and the datetime of the loaded \textsf{homepage2-zone-1} zone-manager file. When the difference is positive, it means that the embedded content (from the \textsf{homepage2-zone-1} zone-manager file) is coming from the future, and when the difference is negative, the embedded content is coming from the past. The black vertical lines indicate the datetime of some of the sparse zone-manager mementos (orange dots in Figure \ref{fig:IA-urims-hp123}). If the embedded resource does not exist at the same datetime as the main page, the Wayback Machine will attempt to find the closest available memento. In the figure, when we see a sharp increase in the difference, that indicates that the next memento is now closer than the previous memento, and the embedded content will be coming from the future rather than the past. We see that on October 8, 2015 there is a spike up to 90 days of difference. Around that day, mementos would be replaying \textsf{homepage2-zone-1} content from July 10, 2015 and then January 6, 2016, corresponding to the nearest zone-manager memento.

\begin{figure*}[ht]
\centering
\begin{subfigure}{0.3\textwidth}
  \centering
  \includegraphics[width=0.99\linewidth]{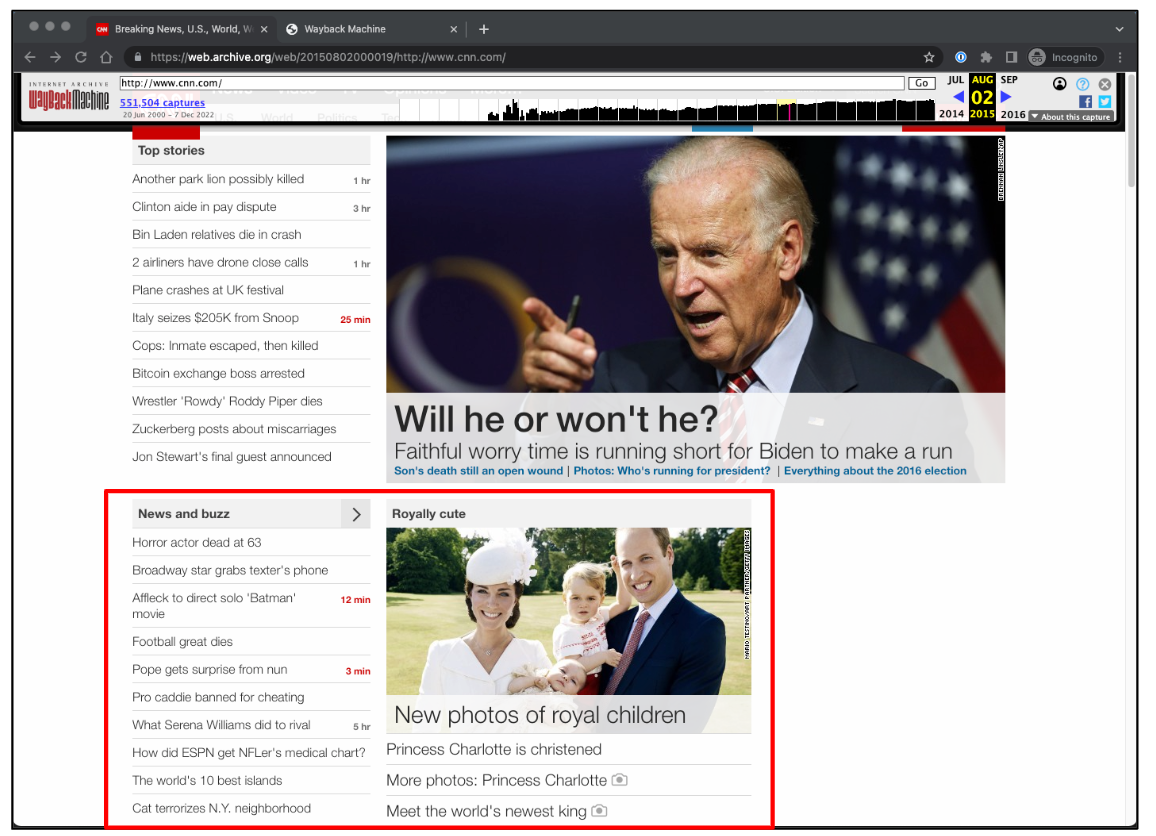}
  \caption{August 2, 2015 (\urim{20150802000019})}
\label{fig:cnn-2015aug02}
\end{subfigure}
\hspace{10pt}
\begin{subfigure}{0.3\textwidth}
  \centering
  \includegraphics[width=0.99\linewidth]{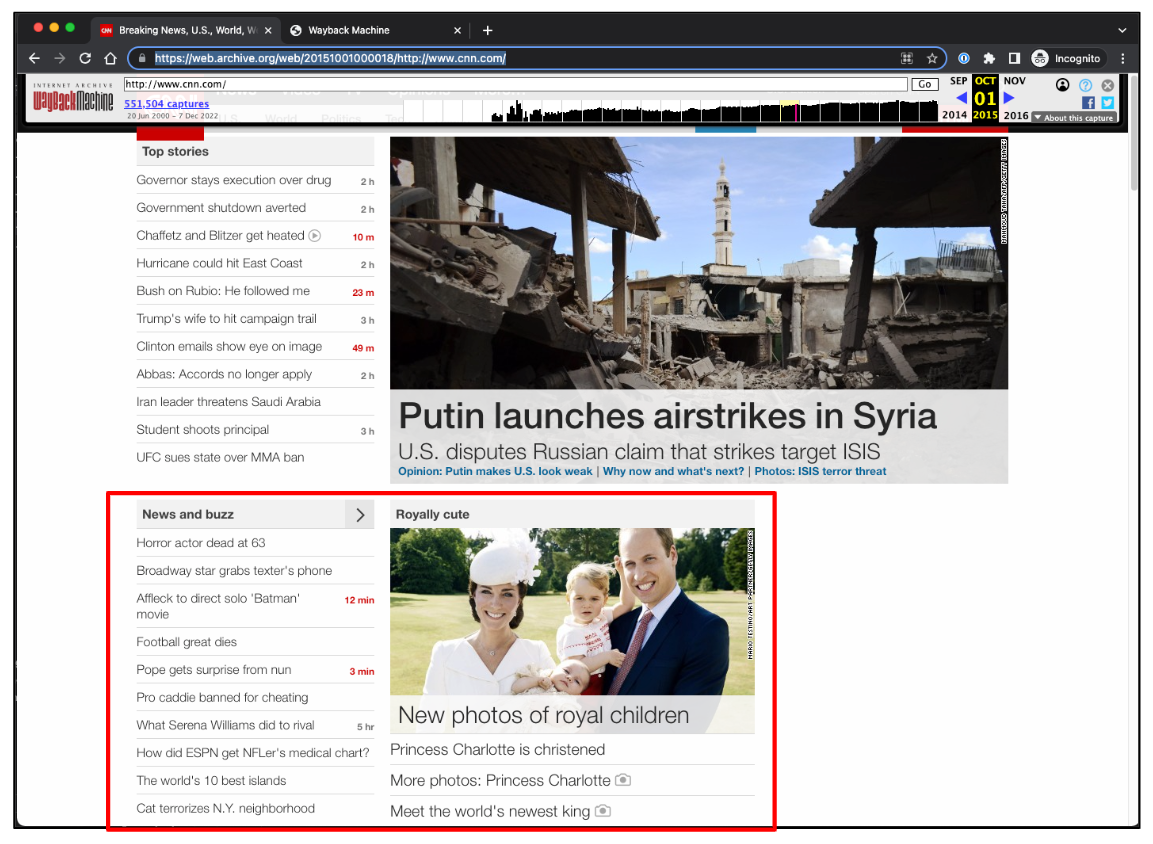}
  \caption{October 1, 2015 (\urim{20151001000018})}
\label{fig:cnn-2015oct01}
\end{subfigure}%
\hspace{10pt}
\begin{subfigure}{0.3\textwidth}
  \centering
  \includegraphics[width=0.99\linewidth]{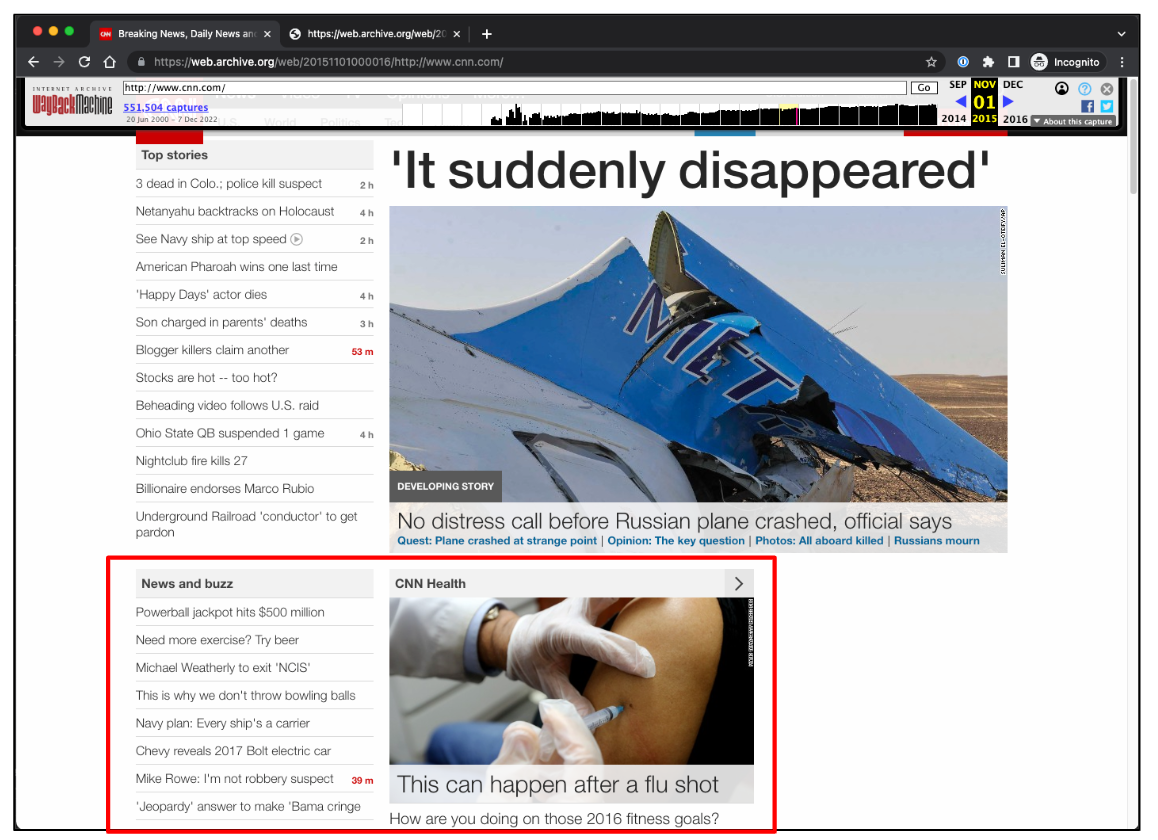}
\caption{November 1, 2015 (\urim{20151101000016})}
\label{fig:cnn-2015nov01}
\end{subfigure}%
\caption{Examples of \textsf{homepage2-zone-1} temporal violations in Wayback Machine replays due to sparsely archived zone-manager files in 2015. Content in (a) and (b) comes from the past, July 10, 2015 (\urim{20150710001845}), and content in (c) comes from the future, January 6, 2016 (\urim{20160106233405}).}
\label{fig:homepage2-zone-1}
\end{figure*}

Figure \ref{fig:homepage2-zone-1} shows some examples of temporal violations with \textsf{homepage2-zone-1} content. 
Figure \ref{fig:cnn-2015aug02} shows a CNN.com memento with the datetime of August 2, 2015 and Figure \ref{fig:cnn-2015oct01} a datetime of October 1, 2015. The Hero content is contained in the base HTML page, so there is no temporal violation. However, the remaining content (zones \textsf{homepage2-zone-1} through \textsf{homepage4-zone-5}) comes from July 2015. This same \textsf{homepage2-zone-1} content can also be seen below the Hero content in Figure \ref{fig:cnn-20150917}.
Figure \ref{fig:cnn-2015nov01} shows an example of content being populated from the future, with the \textsf{homepage2-zone-1} content coming from January 6, 2016. 
This example corresponds to the sharp peak in Figure \ref{fig:IA-hp2-diff}, which occurs on October 8, 2015. Before this point, the mementos replay with the July 10, 2015 \textsf{homepage2-zone-1} content, and after this point, the mementos replay with the January 6, 2016 content -- until a new memento for the \textsf{homepage2-zone-1} content is available on January 7, 2016. 

We computed the number of mementos with various levels of temporal violation in zone \textsf{homepage2-zone-1} (and based on Figure \ref{fig:IA-urims-hp123}, these values will be similar for content from \textsf{homepage3-zone-1}). In Table \ref{tab:threshold}, we consider the impact of employing a temporal range threshold for loading the \textsf{homepage2-zone-1} content during three time periods: April 24 -- May 23, 2015 (no mementos), May 23, 2015 -- July 21, 2016 (sparse mementos) and July 21, 2016 -- September 23, 2022 (dense mementos). 
\begin{table}
  \caption{Mementos Affected With Various Temporal Thresholds for Loading \textsf{homepage2-zone-1} Content from the Internet Archive}
  \label{tab:threshold}
  \begin{tabular}{crrrrr}
    \toprule
     & \multicolumn{2}{c}{Total} & & \multicolumn{2}{c}{Affected}\\
    Date Range&Mementos&Days&Threshold&Mementos&Days\\
    \midrule
    Apr 24, 2015--May 23, 2015& 788 & 29 & 1 hr & 786  &  29 \\
     & & & 2 hrs & 785  &  29 \\
    & & & 6 hrs &  780  &  28 \\
    (none) & & & 24 hrs &  754  &  27 \\
    & & & 48 hrs &  724  &  26 \\
    \midrule
    May 23, 2015--Jul 21, 2016 & 17,017  &  425 & 1 hr & 16,863  &  425\\
     & & & 2 hrs & 16,738  &  425\\
    & & & 6 hrs  & 16,244  &  410\\
    (sparse)& & & 24 hrs & 15,142  &  370\\ 
    & & & 48 hrs &  14,178  &  342\\ 
    \midrule
    Jul 21, 2016--Sep 23, 2022 & 154,917  &  2256 & 1 hr & 95,009  &  1826\\  
    & & & 2 hrs & 67,992  &  1425\\
    &  & & 6 hrs & 13,893  &  190 \\
    (dense) & & & 24 hrs & 1812  &  37\\
    & & & 48 hrs & 526  &  8\\
  \bottomrule
\end{tabular}
\end{table}
In other words, we consider in how many CNN.com mementos (and over how many days) would there be no content in the second (and third) level zones if zone-manager files with temporal differences greater than these thresholds were not allowed to be loaded. We can see that because of the increase in the number of zone-manager mementos after July 21, 2016, the impact of the temporal violation is reduced. However, in the 450 days before then, there would be significant impact.

Temporal violations in the replay of www.cnn.com are even more striking when they occur in the \textsf{homepage1-zone-1} Hero zone as was shown in Figure \ref{fig:cnn-hero-temporal}. Since the CNN.com main page does not contain the datetime, these violations can be difficult to detect since the HTML does not provide a contrasting datetime value to that of the archive. 
We performed some analysis for the \textsf{homepage1-zone-1} content, but since these files are not always requested (as described in Section \ref{sec:construction}), we had to download the www.cnn.com HTML to determine the presence of the \texttt{id="homepage1-zone-1"} attribute before we could determine when this zone is loaded via CSR and would have the potential for a temporal violation. 

\begin{figure}[t]
  \centering
  \includegraphics[width=\linewidth, trim=20 0 70 0, clip]{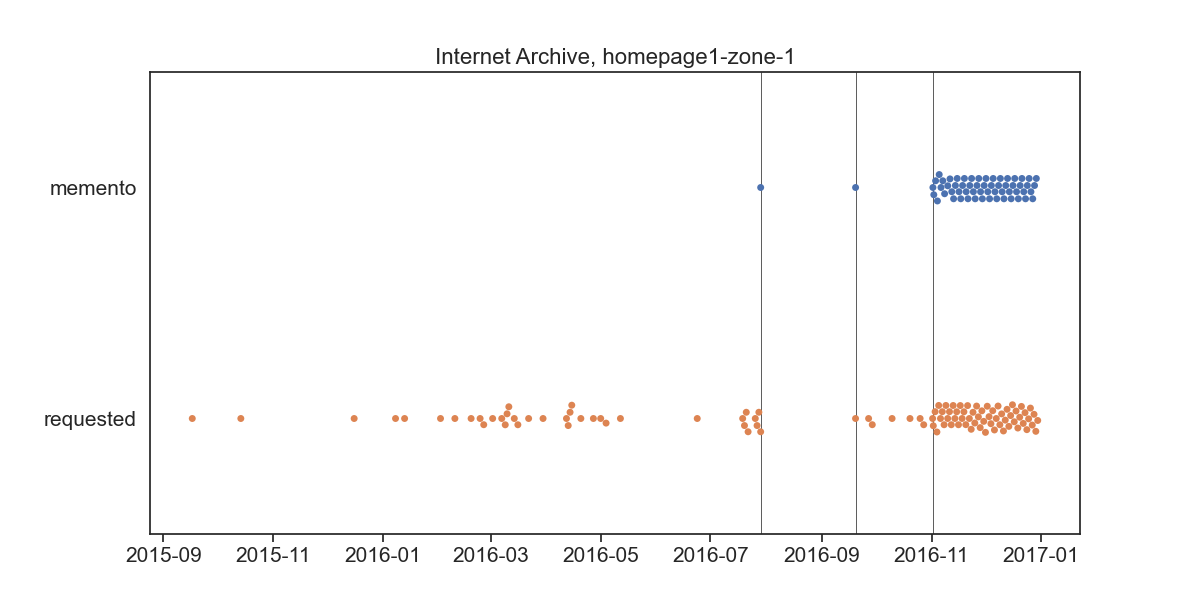}
  \caption{Timeline of mementos and requests from the Internet Archive for the homepage1-zone-1 zone-manager files between September 2015 - December 2016.}
\label{fig:homepage1-zone-mementos-IA}
\end{figure}

We have noted that the earliest memento for the \textsf{homepage1-zone-1} zone-manager is July 29, 2016 (\href{https://web.archive.org/web/20160729003156/https://www.cnn.com/data/ocs/section/index.html:homepage1-zone-1/views/zones/common/zone-manager.html}{20160729003156}). Any CNN.com main page memento before this that requests \textsf{homepage1-zone-1} via CSR will have a temporal violation in the Hero story. 
Figure \ref{fig:homepage1-zone-mementos-IA} shows the timeline of \textsf{homepage1-zone-1} mementos and a sample of CNN.com main pages that request that zone via CSR.  
As with the other charts of this type, a dot indicates that at least one memento was present (or request made) on that day. The vertical lines align with the first three mementos available for the zone. All of the ``requested'' dots before the first line represent a memento with a temporal violation, and there are several instances between the second and third lines. It is important to note that the ``requested'' region does not indicate all days when \textsf{homepage1-zone-1} was requested via CSR, but only the days that we sampled. 
Note that once \textsf{homepage1-zone-1} is regularly requested via CSR in November 2016, the presence of \textsf{homepage1-zone-1} mementos increases.

\begin{figure}[t]
  \centering
  \includegraphics[width=\linewidth]{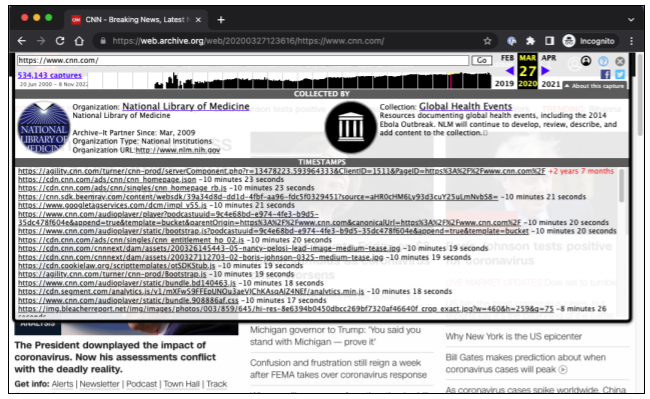}
  \caption{``About this capture'' panel for CNN.com (\urim{20200327123616}), showing 15 of over 200 embedded resources.}
\label{fig:about-this-capture}
\end{figure}

In 2017, the Internet Archive added an ``About this capture'' service to the archival banner \cite{ia:about-this-capture}.  Clicking on the link in the banner will produce a listing of all the resources (e.g., images, stylesheets) used to replay the page and show their associated datetimes and whether or not they were crawled in the future or past, relative to the root HTML page (Figure \ref{fig:about-this-capture}).  However, the display can be challenging for users: as noted in Section \ref{sec:background}, not all temporal spreads are violations \cite{ht15:as-presented}, and the GUI does not scale well for pages with hundreds of embedded resources.  Regardless of these problems, JSON files are currently excluded from the ``About this capture'' display, so the temporal violations in CNN.com happen silently. However, even if JSON files were included, for CNN.com in particular, the typical user would not know the importance of files named \texttt{zone-manager.izl}.

\subsection{Archive-It}

Public resources in Archive-It are made available in the Internet Archive's Wayback Machine, but individual Archive-It collections do not have access to resources outside their collection's holdings.
Because Archive-It collection mementos are limited to resources from that specific collection, there could be even larger gaps between the base page datetime and the content JSON datetime.  

\begin{figure}[ht]
\begin{subfigure}{0.99\linewidth}
\centering
  \includegraphics[width=\linewidth,trim=0 20 70 0, clip]{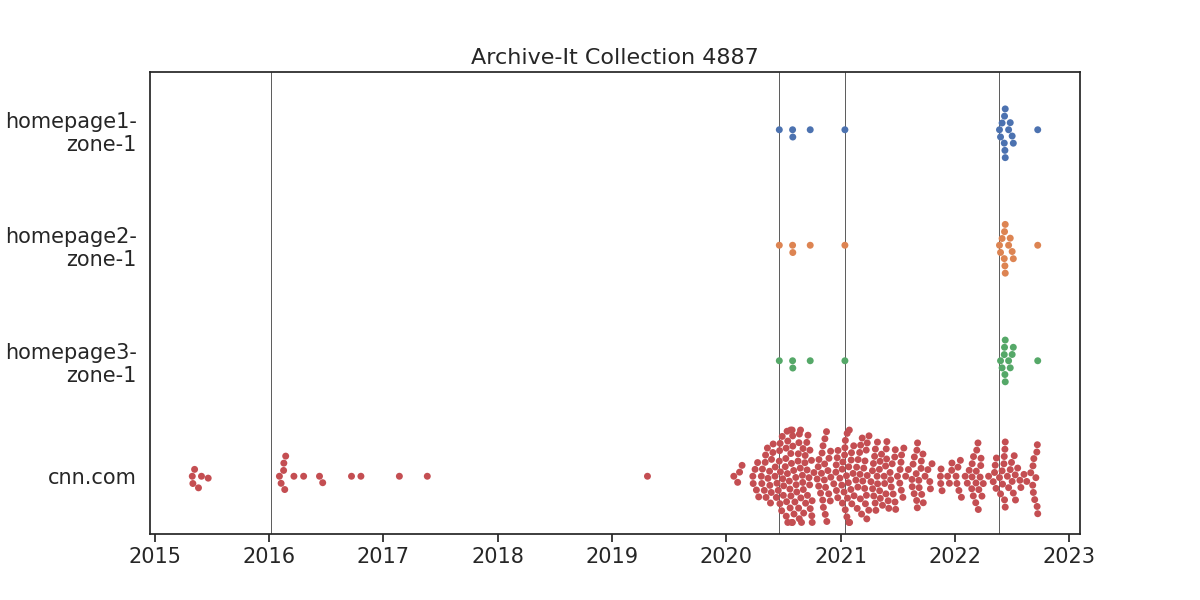}
  \caption{Timeline of mementos for the top three zones and CNN.com.}
\label{fig:zone-mementos-AIT-4887}
\end{subfigure}
\begin{subfigure}{0.99\linewidth}
  \centering
  \includegraphics[width=\linewidth,trim=0 20 70 0, clip]{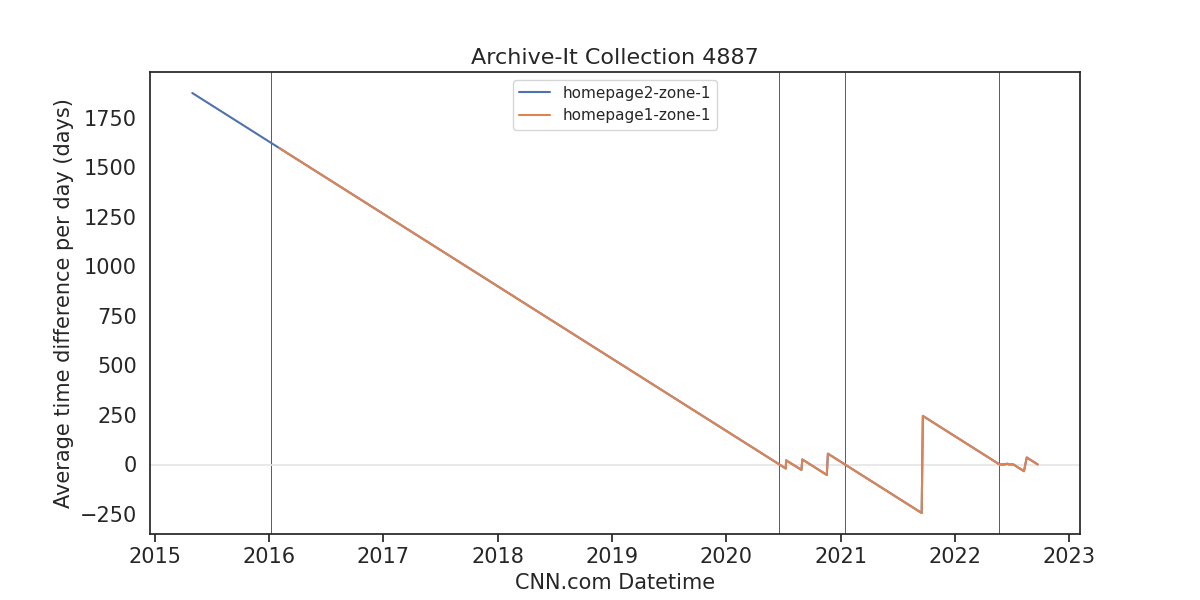}
  \caption{Average difference in days between the CNN.com datetime and the closest memento of the \textsf{homepage1-zone-1} and \textsf{homepage2-zone-1} zone-manager files}
\label{fig:AIT-4887-hp2-diff}
\end{subfigure}
\caption{Analysis of CNN.com Mementos in Archive-It Collection 4887} \label{fig:ait-4887}
\end{figure}

The National Library of Medicine's Global Health Events collection\footnote{\url{https://archive-it.org/collections/4887}} (collection 4887) at Archive-It is an example of a large (22,000 seeds) and long-running (since October 2014) collection. It is especially valuable due to its focused curation of webpages related to the COVID-19 pandemic.  
Figure \ref{fig:ait-4887} shows the mementos and difference charts for this collection. This collection has 600 mementos with status code 200 OK between May 2015-December 2022. The vertical lines in both charts mark the first time that \textsf{homepage1-zone-1} is requested via CSR (in 2016) and first days when the zone-manager mementos are captured after a gap.
Figure \ref{fig:zone-mementos-AIT-4887} shows a plot of the mementos available in the collection, and Figure \ref{fig:ait-4887} shows the corresponding difference between the datetime of the CNN.com main page and the \textsf{homepage1-zone-1} and \textsf{homepage2-zone-1} zone-manager files.  The lines overlap because mementos of the zone-manager files were captured at the same time.
Note that the difference here is shown in days rather than hours. We note that although CNN.com was captured by this collection starting in 2015, the first memento of a zone-manager file was not captured until June 2020.

\begin{figure}[ht]
\begin{subfigure}{\linewidth}
\centering
  \includegraphics[width=\linewidth,trim=0 20 70 0, clip]{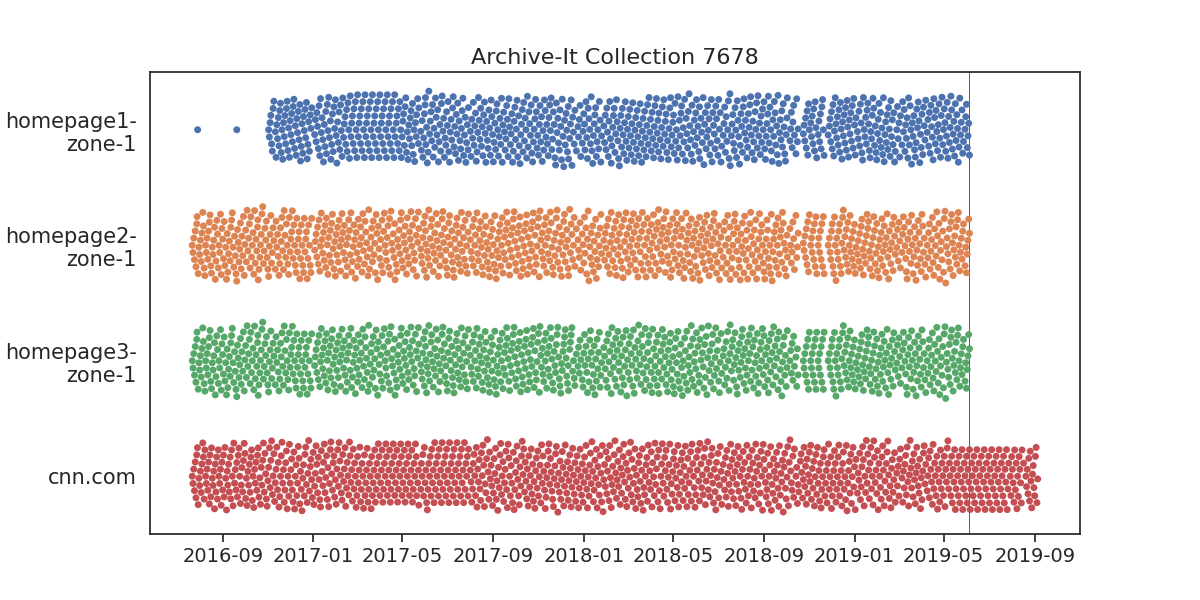}
  \caption{Timeline of mementos for the top three zones and CNN.com.}
\label{fig:zone-mementos-AIT-7678}
\end{subfigure}
\begin{subfigure}{\linewidth}
  \centering
  \includegraphics[width=\linewidth,trim=0 20 70 0, clip]{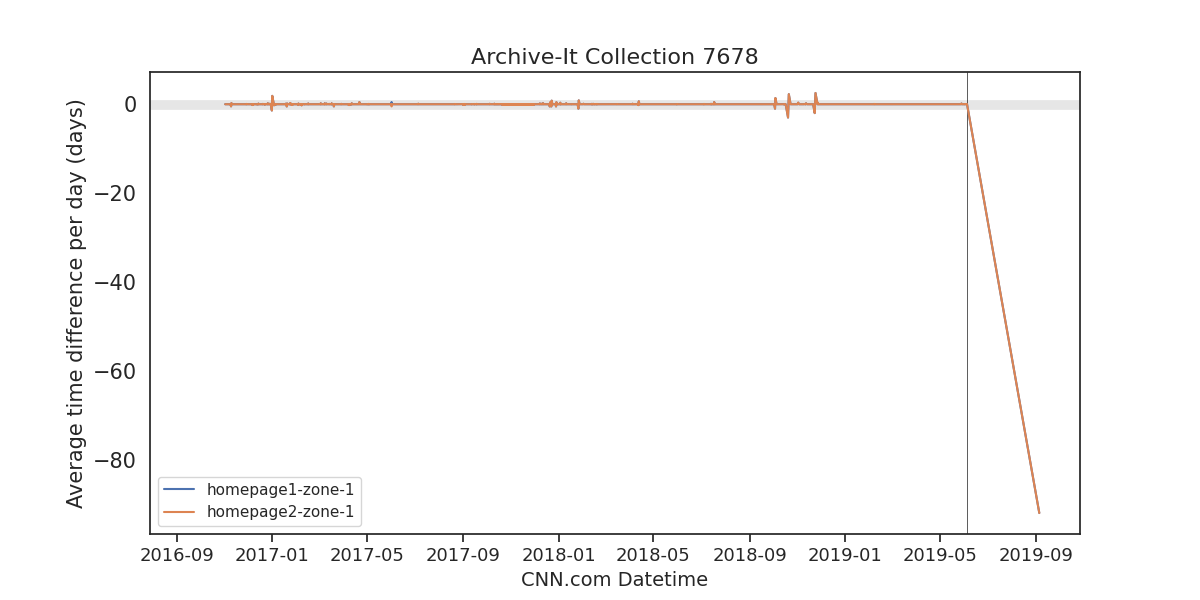}
  \caption{Average difference in days between the CNN.com datetime and the closest memento of the \textsf{homepage1-zone-1} and \textsf{homepage2-zone-1} zone-manager files}
\label{fig:AIT-7678-hp2-diff}
\end{subfigure}
\caption{Analysis of CNN.com Mementos in Archive-It Collection 7678} \label{fig:ait-7678}
\end{figure}

However, there are other Archive-It collections (owned by different organizations) that do have a substantial number of CNN.com mementos and associated resources. For instance, collection 7678\footnote{\url{https://archive-it.org/collections/7678}} was created by Mark Graham, Director of the Wayback Machine, and has a single seed: \texttt{http://www.cnn.com}. This collection has 2619 mementos of the main page of CNN.com between July 2016 and September 2019, along with almost 13,000 mementos of the various zone-manager files.  Figure \ref{fig:zone-mementos-AIT-7678} shows a plot of the mementos available for CNN.com and the top zone-manager files (zone-1) in Archive-It collection 7678. 
The corresponding difference line (in days) is shown in Figure \ref{fig:AIT-7678-hp2-diff}. 
The vertical line shows the last \textsf{homepage2-zone-1} zone-manager memento in that collection (June 4, 2019), although the CNN.com main page was continued to be captured until September 2019. The shaded area around 0 on the y-axis shows a threshold of 1 day. We can see that this collection provides a set of high quality captures of CNN.com, with most of the differences being contained in the 1-day range.
If other Archive-It collections were allowed to borrow zone-manager files from this collection, playback would be greatly improved.

\section{Discussion}

This analysis points several items for further discussion. First, in the short term, what should web archives like the Internet Archive do about these violative mementos? We have seen that simply displaying the amount of time difference of various embedded resources is not effective for pages, like CNN.com, that potentially have hundreds of embedded resources. Web archives could consider the amount of temporal difference in redirected resources before requesting mementos from different datetimes. This would likely require specialized code and analysis of affected pages (similar to what we have done here), because as we have mentioned, for many embedded resources, like images, it may not matter if the temporal difference is minutes vs. days vs. months. For CNN.com, archives could ensure that resources with extensions of \texttt{.html}, \texttt{.izl.json}, or \texttt{.izl} are not loaded with large temporal differences. That would address the temporal violations in the past, however, nothing prevents CNN.com from changing the filename extension at some point in the future. In addition, not including those resources would result in mementos that replay with significant amounts of missing content. Because of its abundance of archived material, the Internet Archive has another option for mementos with temporal violations in the Hero zone - silently delete those mementos. For each memento that we have identified with a temporal violation in \textsf{homepage1-zone-1}, there is another memento where that content is included in the HTML, resulting in no temporal violation. This only holds for mementos captured before November 1, 2016. For Archive-It, might it be possible for all collections that have mementos of CNN.com between July 2016 - June 2019 to borrow zone-manager resources from collection 7678 (or from general Internet Archive holdings)? This would likely improve the replay of CNN.com mementos in many Archive-It collections. 

Further, what should web archives be doing to avoid this going forward? This problem is not just limited to social media sites and CNN.com. We have observed (but not fully measured) similar behavior in the main web page for the \emph{Atlanta Journal-Constitution}, AJC.com. Given that CNN.com moved to client-side rendering in 2015, this is not a new web design technique. There are certainly many webpages that are affected by similar issues, they just are not prominent enough to be detected.  These webpages should be targeted for archiving using browser-based crawlers, similar to how the Internet Archive employs browser-based crawling for certain webpages and for those that are requested to be crawled via the ``Save Page Now'' feature. 

In addition to the CNN.com specific heuristic of looking for the absence of \texttt{id="homepage1-zone-1"} in the HTML to determine when the Hero content was loaded via CSR, we also identified another heuristic to detect HTML that is largely only providing a client-side rendering template. We found when CNN.com requests all zones via CSR, the amount of likely content text (words starting with lowercase characters) in the base HTML is 15 words or less. This is after after all HTML tags, \texttt{script}, and \texttt{style} content has been removed. This heuristic works for CNN.com, but future study could be done to see how widely applicable this or a similar heuristic could be. Then for those pages detected to be heavily built with client-side rendering, those should be flagged for crawling with a browser-based crawler. 

\section{Conclusions}

As a result of client-side rendering of HTML pages, where JSON resulting from API calls is used to build out the DOM, 
it can be difficult to archive and replay such pages using non-browser-based web archiving tools.  Heritrix is one such tool, and it is optimized for speed and site-level crawls, whereas browser-based tools are more complete and operate at a much slower speed.  It is well-known that conventional tools like Heritrix often fail to adequately archive social media sites, but we have shown that there are implications for news sites, such as CNN.com, as well.  As a result, a page archived at one datetime can replay news headlines, loaded from a violative JSON response, that are far in the future or the past relative to the datetime of the main HTML page.  Since the violative JSON is not directly observable in the page in the same way an image is, for example, detecting the temporal violations can be difficult.  This is problematic, as the contents of web archives have been used for evidentiary purposes and large differences between what was captured and what is replayed should be easily apparent to the user.

For CNN.com's main page, we have measured the temporal violations in Hero and second and third level content. During the 453-day period of no and sparse archiving of zone-manager files, there are significant temporal violations in the second and third level content, up to 90 days difference. We also found instances of significant temporal violation in Hero content, but this could be mitigated by redacting those particular mementos, because other mementos exist on those days that have the Hero content delivered in the HTML.   
The impact of this is not limited to CNN.com, but could affect any often-updated webpage that uses client-side rendering. Maintaining page coherency, that is, making sure the JSON responses are archived at the same time as the initial HTML response, requires browser-based crawling.  Unfortunately, browser-based crawlers are not yet optimized to be fast enough to crawl entire sites at scale, so we might need methods to identify which pages require browser-based crawling and which are suitable for faster tools such as Heritrix.  

\bibliographystyle{ACM-Reference-Format}
\bibliography{refs}

\end{document}